\documentclass[aps,prd,10pt,notitlepage,nofootinbib]{revtex4-1}

\usepackage[utf8]{inputenc}
\usepackage{amsmath,amssymb,amsfonts}
\usepackage{newtxtext,newtxmath}

\usepackage{bbold}
\usepackage{bm}
\usepackage{graphicx}
\usepackage[usenames,dvipsnames]{xcolor}
\usepackage{color}
\usepackage[colorlinks=true,linkcolor=blue,urlcolor=blue,citecolor=blue]{hyperref}

\usepackage{slashed}
\usepackage[english]{babel}
\usepackage{dcolumn}
\usepackage{pifont}
\usepackage{dsfont,mathrsfs}
\usepackage{cancel}
\usepackage{bigints}
\usepackage{accents}
\usepackage{soul}
\usepackage{multirow}

\usepackage{amsmath, amssymb}
\usepackage{bbold}
\usepackage{makecell}
\usepackage{simpler-wick}

\usepackage{orcidlink} 

\newcommand{\nn}{\nonumber}

\newcommand{\ensembleaverage}[1]{\left\langle#1\right\rangle}

\newcommand{\MB}[1]{\left|#1\right|}
\newcommand{\mb}[1]{|#1|}
\newcommand{\FB}[1]{\left(#1\right)}
\newcommand{\fb}[1]{(#1)}
\newcommand{\SB}[1]{\left\{#1\right\}}
\newcommand{\TB}[1]{\left[#1\right]}

\newcommand{\scrM}{\mathscr{M}}

\newcommand{\munu}{{\mu\nu}}

\newcommand{\alphabeta}{{\alpha\beta}}
\newcommand{\mnab}{{\mu\nu\alpha\beta}}

\newcommand{\IM}{\text{Im}}
\newcommand{\RE}{\text{Re}}
\newcommand{\Tr}{\varmathbb{Tr}}

\newcommand{\psibar}{\overline{\psi}}
\newcommand{\qbar}{\bar{q}}

\newcommand{\del}{\partial}
\newcommand{\identity}{\mathds{1}}

\newcommand{\wpr}{\omega^r_p}
\newcommand{\wopr}{\omega^r_{0p}}
\newcommand{\wps}{\omega^s_p}
\newcommand{\wkr}{\omega^r_k}
\newcommand{\wks}{\omega^s_k}
\newcommand{\fkr}{f_k^r}
\newcommand{\fks}{f_k^s}
\newcommand{\fpr}{f_p^r}
\newcommand{\fps}{f_p^s}

\newcommand{\wk}{\omega_k}

\newcommand{\mans}{\mathfrak{s}}

\newcommand{\rs}{\sqrt{\mans}}

\newcommand{\mant}{\mathfrak{t}}
\newcommand{\manu}{\mathfrak{u}}

\allowdisplaybreaks

\begin{document}
	\title{Shear and bulk viscous coefficients of a hot and chirally imbalanced quark matter using NJL model}

\author{Snigdha Ghosh\orcidlink{0000-0002-2496-2007}$^{a}$}
\email{snigdha.ghosh@bangla.gov.in}
\email{snigdha.physics@gmail.com}
\thanks{Corresponding Author}
\author{Nilanjan Chaudhuri\orcidlink{0000-0002-7776-3503}$^{b,d}$}
\email{n.chaudhri@vecc.gov.in}
\email{nilanjan.vecc@gmail.com}
%
\author{Pradip Roy\orcidlink{0009-0002-7233-4408}$^{c,d}$}
\email{pradipk.roy@saha.ac.in}
\author{Sourav Sarkar\orcidlink{0000-0002-2952-3767}$^{b,d}$}
\email{sourav@vecc.gov.in}
\affiliation{$^a$Government General Degree College Kharagpur-II, Paschim Medinipur - 721149, West Bengal, India}	
\affiliation{$^b$Variable Energy Cyclotron Centre, 1/AF Bidhannagar, Kolkata - 700064, India}
\affiliation{$^c$Saha Institute of Nuclear Physics, 1/AF Bidhannagar, Kolkata - 700064, India}
\affiliation{$^d$Homi Bhabha National Institute, Training School Complex, Anushaktinagar, Mumbai - 400085, India}
	
\begin{abstract}
The shear $\eta$ and bulk $\zeta$ viscous coefficients have been calculated in a hot and chirally asymmetric quark matter quantified in terms of a chiral chemical potential (CCP) using the two-flavor Nambu-Jona--Lasinio (NJL) model. This is done by employing the one-loop Green-Kubo formalism where the viscous coefficients have been extracted from the long-wavelength limit of the in-medium spectral function corresponding to the energy momentum tensor (EMT) current correlator calculated using the real time formalism of finite temperature field theory. The momentum dependent thermal width of the quark/antiquark that enters into the expression of the viscosities as a dynamical input containing interactions, has been obtained from the $2\to2$ scattering processes mediated via the collective mesonic modes in scalar and pseudoscalar channels encoded in respective in-medium polarization functions having explicit temperature and CCP dependence. Several thermodynamic quantities such as pressure, energy density, entropy density $(s)$, specific heat and isentropic speed of sound have also been calculated at finite CCP. The temperature and CCP dependence of the viscosity to entropy density ratios $\eta/s$ and $\zeta/s$ have also been studied.
\end{abstract}

\maketitle

\section{Introduction}

The study of quantum chromodynamics (QCD) matter under extreme conditions of temperature and/or baryon density is one of the main objectives of relativistic heavy ion collision (HIC) experiments at RHIC and LHC. It is well known that there exists an infinite number of energy-degenerate different vacuum configurations of QCD at zero and low temperatures which are characterized by topologically nontrivial gauge configurations with a nonzero winding number~\cite{Shifman:1988zk} called instantons. Transitions between two different vacua occur by tunneling through a potential barrier with a height of the order of the QCD scale $\Lambda_{QCD}$~\cite{Belavin:1975fg,tHooft:1976rip,tHooft:1976snw}. Another type of gluon configuration called sphalerons are expected to be produced copiously at high temperatures for example, in the quark gluon plasma (QGP) phase of HICs~\cite{Manton:1983nd,Klinkhamer:1984di} which can enhance the transition rate
between different vacua~\cite{Kuzmin:1985mm,Arnold:1987mh,Khlebnikov:1988sr,Arnold:1987zg}. This mechanism can flip the helicities of quarks leading to the breaking of parity ($ P $) and charge-parity ($ CP $) symmetries by creating an asymmetry between left and right handed quarks via the axial anomaly of QCD~\cite{Adler:1969gk,Bell:1969ts}. This leads to local chiral imbalance as there is no direct observation	of global violation of $P$ and $CP$ in QCD~\cite{Adler:1969gk,Bell:1969ts,McLerran:1990de,Moore:2010jd}. The chiral imbalance is characterized by means of a chiral chemical potential (CCP) which represents the difference between the number of right and left-handed quarks/antiquarks. Chiral imbalance in the presence of high magnetic fields produced in noncentral HICs leads to exotic phenomena such as chiral magnetic effect  (CME)~\cite{Fukushima:2008xe,Kharzeev:2007jp,Kharzeev:2009pj,Bali:2011qj} which is presently being intensively looked for in HICs~\cite{STAR:2021mii}.
In addition to the CME, intense efforts have also been made in the recent past to study the phase structure ~\cite{Ruggieri:2016lrn,Ruggieri:2016asg,Ruggieri:2020qtq},  microscopic transport phenomena ~\cite{Vilenkin:1979ui,Vilenkin:1980fu,Fukushima:2008xe,Son:2009tf}, collective oscillations ~\cite{Akamatsu:2013pjd,Carignano:2018thu,Carignano:2021mrn}, fermion damping rate~\cite{Carignano:2019ivp} and collisional energy loss of fermions~\cite{Carignano:2021mrn} as well as properties of electromagnetic spectral function\cite{Ghosh:2022xbf} in chirally imbalanced medium.

The comparison of the experimentally measured elliptic flow $v_2$ in peripheral HICs with viscous hydrodynamics has indicated the production of a nearly perfect fluid with a very small shear viscosity to entropy density ratio $\eta/s$. In particular, the elliptic flow parameter exhibits a strong dependence on the $\eta/s$ of the dissipative quark gluon matter formed in the collision.
There are considerable number of works aiming to calculate the viscous coefficients of hot and/or dense QCD medium using NJL-like models~\cite{Zhuang:1995uf,Rehberg:1996vd,Iwasaki:2007iv,Alberico:2007fu,Sasaki:2008um,Marty:2013ita,Lang:2013lla,Lang:2015nca,Ghosh:2015mda,Harutyunyan:2017ttz,Zhang:2021xib}. In Refs.~\cite{Zhuang:1995uf,Rehberg:1996vd}, shear viscosity of hot quark matter has been calculated within two/three-flavor NJL model using an elementary kinetic theory expression of $\eta$. In Refs.~\cite{Iwasaki:2007iv,Alberico:2007fu}, the authors used Kubo formula to obtain $\eta$ at finite temperature using NJL model in which a different modeling of the quark spectral function was used. Refs.~\cite{Sasaki:2008um,Marty:2013ita} use a two/three flavor NJL and/or dynamical quasi particle model to calculate the shear as well as bulk viscous coefficients at finite temperature and density employing the relaxation time approximation (RTA) of the relativistic kinetic theory. The authors of Refs.~\cite{Lang:2013lla,Lang:2015nca,Harutyunyan:2017ttz} used the general Kubo formula to calculate $\eta$ within NJL model in a hot and dense quark medium, where the full Dirac structure of the thermal quark spectral function including contributions from mesonic fluctuations are taken into consideration. However, to the best of our knowledge, we have not come across any calculation of the viscosity of  quark matter at finite temperature and CCP using NJL-like models.

In this work we aim to calculate the shear $\eta$ and bulk $\zeta$ viscous coefficients of hot and chirally imbalanced quark matter matter employing the one-loop Green-Kubo formalism. For this we will first evaluate within the mean field approximation (MFA) of the Nambu-Jona--Lasinio (NJL) model, the in-medium spectral function corresponding to the energy momentum tensor (EMT) current correlator using the real time formalism (RTF) of finite temperature field theory at finite CCP. The viscous coefficients will then be extracted from the long-wavelength (low frequency) limit of the in-medium spectral function. In the expression of the viscosities,  the thermal width of the quark/antiquark having momentum, temperature and CCP dependence, enters as a dynamical input containing interactions owing to the spectral-broadening of the quarks in the thermal medium; the viscosities of a noninteracting system is otherwise infinite like an ideas gas. The quark/antiquark thermal width will be obtained considering $2\to2$ scattering processes among the quarks/antiquarks mediated via the collective mesonic modes in scalar ($\pi$) and pseudoscalar ($\sigma$) chanels encoded in their respective in-medium polarization functions having explicit temperature and CCP dependence. The mesonic polarization functions will also be calculated employing the RTF at finite CCP. Some thermodynamic quantities such as pressure ($P$), energy density ($\varepsilon$), entropy density ($s$), specific heat ($C_V$) and isentropic speed of sound $(c_s)$ will also be calculated at finite CCP in the NJL model. We will also study the temperature and CCP dependence of the viscosity to entropy density ratios $\eta/s$ and $\zeta/s$.

The article is organized as follows. In Sec.~\ref{sec.viscosity} the main formalism of calculating viscous coefficients in NJL model has been discussed. Next in Sec.~\ref{sec.width}, the estimation of quark thermal width is outlined followed by the calculation of thermodynamic quantities including the speed of sound in Sec.~\ref{sec.thermodynamics}. After that, we present the numerical results in Sec.~\ref{sec.results} and finally summarize and conclude in Sec.~\ref{sec.summary}.

\textit{Notations}: The metric tensor has the signature $\text{diag}(1,-1,-1,-1)$. Four-vector are denoted by capital letters as $P\equiv(p^0,\bm{p})$ and $p=|\bm{p}|$ represents the magnitude of the three-vector.

\section{Viscous Coefficients at finite CCP from NJL Model} \label{sec.viscosity}
The standard expression of the Lagrangian (density) in the MFA for two-flavor NJL model at nonzero CCP $\mu_5>0$ is given by  
\begin{eqnarray}
	\mathscr{L}_\text{NJL}^\text{MFA} = \psibar( \frac{1}{2}\gamma^\mu i\overleftrightarrow{\del_\mu}-M + \mu_5\gamma^0\gamma^5)\psi - \frac{(M-m)^2}{4G} \label{lagrangian.mfa}
\end{eqnarray} 
where, $\psi = \FB{ \begin{array}{c} u \\ d \end{array} }$ is the quark isospin doublet, the notation $A\overleftrightarrow{\del_\mu}B = A(\del_\mu B) - (\del_\mu A)B$, $G$ is the NJL scalar coupling, $m$ is the current quark mass, and $M$ is the constituent quark mass obtained from the gap equation
\begin{eqnarray}
	M = m -2G\ensembleaverage{\psibar\psi} \label{gap}
\end{eqnarray}
in which $\ensembleaverage{\psibar\psi}$ is the chiral condensate. In the MFA, $\ensembleaverage{\psibar\psi}$ is given by
\begin{eqnarray}
	\ensembleaverage{\psibar \psi} = \RE~i\int\!\!\frac{d^4P}{(2\pi)^4} \Tr \TB{S^{11}(P)}, \label{condensate}
\end{eqnarray}
where the trace $\Tr[...]$ has to be taken over Dirac as well as the flavor and color spaces, and $S^{11}(P)$ is the $11$ component of the thermal quark propagator in the RTF defined with respect to a symmetric Keldysh-Schwinger contour in the complex time plane~\cite{Bellac:2011kqa,Mallik:2016anp} and given by the following expression~\cite{Ghosh:2022xbf}
\begin{eqnarray}
	S^{11}(P) = \mathcal{D}(P) \sum_{r\in\{\pm\}}^{}\frac{1}{4pr\mu_5}\TB{\frac{-1}{(p^0)^2-(\wpr)^2+i\epsilon}-2\pi i f(\wpr) \delta\SB{(p^0)^2-(\wpr)^2}}. \label{S11}
\end{eqnarray}
In Eq.~\eqref{S11}, $\wpr = \sqrt{(p+r\mu_5)^2+M^2}$ is the single particle dispersion relation of the quark propagating with helicity $r$ and three-momentum $\bm{p}$ in presence of CCP, $f(x) = \dfrac{1}{e^{x/T}+1}$ is the Fermi-Dirac thermal distribution function at temperature $T$, and $\mathcal{D}(P)$ is given by
\begin{eqnarray}
	\mathcal{D}(P) = \sum_{j \in \{\pm\}} \mathscr{P}_j \TB{ P_{-j}^2\cancel{P}_j - M^2 \cancel{P}_{-j} + M (P_j\cdot P_{-j}-M^2) - i M \sigma_\munu P_j^\mu P_{-j}^\nu} 
	\otimes\identity_\text{flavor}\otimes\identity_\text{color} \label{D}~,
\end{eqnarray}
where $\mathscr{P}_j = \frac{1}{2}(\identity+j\gamma^5)$ are the chiral projectors, and $P^\mu_j \equiv (p^0 +j\mu_5,\bm{p})$. (Note that the typographical error in the sign of the $\sigma_\munu$-term in the above equation present in our earlier work~\cite{Ghosh:2022xbf} has been corrected). Substituting the quark propagator $S_{11}(P)$ from Eq.~\eqref{S11} into Eq.~\eqref{condensate} and evaluating the $dp^0$ integration we get after simplification,
\begin{eqnarray}
	\ensembleaverage{\psibar\psi} = -N_c N_f M \sum_{r \in \{\pm\}} \int\!\! \frac{d^3p}{(2\pi)^3} \frac{1}{\wpr} \SB{ f_\Lambda(p)-2f(\wpr)}, \label{condensate.2}
\end{eqnarray}
where $N_c=3$ and $N_f=2$ are respectively the number of colors and flavors, and $f_\Lambda(p)=\sqrt{\dfrac{\Lambda^{20}}{\Lambda^{20} + p^{20}}}$ is a smooth three-momentum cutoff regulator that has been introduced by hand to regulate the ultraviolet (UV) divergences appearing from the momentum integrals in which $\Lambda$ is the three-momentum cutoff parameter. The nonrenormalizable NJL model requires a particular regularization scheme to deal with the UV divergences~\cite{Klevansky:1992qe}. 

In order to calculate the viscous coefficients, we will be using Green-Kubo formalism in which the viscosities are related to the long-wavelength limit of the in-medium spectral function $\rho^\mnab$ corresponding to the energy momentum tensor (EMT) current correlator defined as
\begin{eqnarray}
	\rho^\mnab = \lim\limits_{q^0\to0, \bm{q}=\bm{0}} \frac{\del}{\del q^0} \TB{ \tanh\FB{\frac{q^0}{2T}} \IM~ i \int\! d^4X e^{iQ\cdot X} \ensembleaverage{\mathcal{T}_C T^\munu(X)T^\alphabeta(0)}^{11}} \label{sp.fn}~,
\end{eqnarray} 
where $T^\munu(X)$ is the local EMT current, $\ensembleaverage{\mathcal{T}_C T^\munu(X)T^\alphabeta(Y)}^{11}$ is the $11$-component of the time-ordered thermal two-point EMT-EMT current correlation function in RTF; the time-ordering $\mathcal{T}_C$ is taken with respect to the symmetric Keldysh-Schwinger complex time contour $C$. From the Lagrangian in Eq.~\eqref{lagrangian.mfa}, symmetric EMT (in its Lorentz indices) can be constructed as 
\begin{eqnarray}
	T^\munu(X) &=& \frac{i}{2} \FB{ \psibar \gamma^\mu \overleftrightarrow{\del^\nu}\psi + \psibar \gamma^\nu \overleftrightarrow{\del^\mu}\psi} - g^\munu \mathscr{L}_\text{NJL}^\text{MFA} \nn \\
	&=& \psibar_a(X) \mathcal{O}_{Xab}^{+\munu}\psi_b(X) - \mathcal{O}_{Xab}^{-\munu}\psibar_a(X) \psi_b(X) + g^\munu \frac{(M-m)^2}{4G} \label{EMT}
\end{eqnarray} 
where each of the index $a=\{a_D,a_c,a_f\}$ and $b=\{b_D,b_c,b_f\}$ represents the set of all the symmetry group indices viz Dirac ($D$), color ($c$) and flavor ($f$) to which the field multiplet $\psi$ belongs and the differential operator $\mathcal{O}_{Xab}^{\pm\munu}$ is given by
\begin{eqnarray}
	\mathcal{O}_{Xab}^{\pm\munu} = \frac{1}{2}\TB{ \gamma^\mu i\del_X^\nu + \gamma^\nu i\del_X^\mu - g^\munu(i\cancel{\del}_X \pm \mu_5\gamma^0\gamma^5 \mp M)}_{ab}. \label{O}
\end{eqnarray}
Using Eq.~\eqref{EMT}, the two-point correlator $\ensembleaverage{\mathcal{T}_C T^\munu(X)T^\alphabeta(Y)}^{11}$ can be written as
\begin{eqnarray}
	\ensembleaverage{\mathcal{T}_C T^\munu(X)T^\alphabeta(Y)}^{11} = \ensembleaverage{\mathcal{T}_C
\Big\{\psibar_a(X) \mathcal{O}_{Xab}^{+\munu}\psi_b(X) - \mathcal{O}_{Xab}^{-\munu}\psibar_a(X) \psi_b(X)\Big\}
\Big\{\psibar_c(Y) \mathcal{O}_{Ycd}^{+\alphabeta}\psi_d(Y) - \mathcal{O}_{Ycd}^{-\alphabeta}\psibar_c(Y) \psi_d(Y)\Big\}}^{11}
\end{eqnarray}
which on applying the Wick's theorem becomes
\begin{eqnarray}
		\ensembleaverage{\mathcal{T}_C T^\munu(X)T^\alphabeta(Y)}^{11} &=& 
		\wick{ \ensembleaverage{\mathcal{T}_C \c2\psibar_a(X) \mathcal{O}_{Xab}^{+\munu}\c1\psi_b(X)  \c1\psibar_c(Y) \mathcal{O}_{Ycd}^{+\alphabeta}\c2\psi_d(Y) }^{11}  }
	-	\wick{\ensembleaverage{\mathcal{T}_C \c2\psibar_a(X) \mathcal{O}_{Xab}^{+\munu} \c1\psi_b(X)   \mathcal{O}_{Ycd}^{-\alphabeta}\c1\psibar_c(Y) \c2\psi_d(Y) }^{11} }\nn \\
&& - \wick{\ensembleaverage{\mathcal{T}_C \mathcal{O}_{Xab}^{-\munu} \c2 \psibar_a(X) \c1\psi_b(X)  \c1\psibar_c(Y) \mathcal{O}_{Ycd}^{+\alphabeta} \c2\psi_d(Y) }^{11}}
+ \wick{\ensembleaverage{\mathcal{T}_C  \mathcal{O}_{Xab}^{-\munu} \c2\psibar_a(X) \c1\psi_b(X) \mathcal{O}_{Ycd}^{-\alphabeta}\c1\psibar_c(Y) \c2\psi_d(Y) }^{11}} \nn \\
&=& (-)\Big[ \mathcal{O}_{Xab}^{+\munu} S^{11}_{bc}(X-Y) \mathcal{O}_{Ycd}^{+\alphabeta} S^{11}_{da}(Y-X)
- \mathcal{O}_{Xab}^{+\munu} \mathcal{O}_{Ycd}^{-\alphabeta} S^{11}_{bc}(X-Y) S^{11}_{da}(Y-X) \nn \\
&& - S^{11}_{bc}(X-Y) \mathcal{O}_{Xab}^{-\munu} \mathcal{O}_{Ycd}^{+\alphabeta} S^{11}_{da}(Y-X)
+ \mathcal{O}_{Ycd}^{-\alphabeta} S^{11}_{bc}(X-Y) \mathcal{O}_{Xab}^{-\munu} S^{11}_{da}(Y-X) \Big]
\label{corr.1}
\end{eqnarray}
where, an overall negative sign appears due to the anticommuting nature of the Fermionic fields and 
$S^{11}_{ab}(X-Y) = \wick{ \ensembleaverage{\mathcal{T}_C \c1\psi_a(X) \c1\psibar_b(Y) }^{11} }$ is the 11-component of the real time coordinate space thermal quark propagator at finite CCP. $S^{11}(X-Y)$ can be 
Fourier transformed as
\begin{eqnarray}
	S^{11}(X-Y) = \int\!\!\frac{d^4P}{(2\pi)^4} e^{-iP\cdot(X-Y)} \FB{-i S^{11}(P)} \label{S11.XY}
\end{eqnarray}
where the corresponding momentum space propagator $S^{11}(P)$ is given in Eq.~\eqref{S11}. Substituting Eq.~\eqref{S11.XY} into Eq.~\eqref{corr.1} we obtain after some simplification
\begin{eqnarray}
	\ensembleaverage{\mathcal{T}_C T^\munu(X)T^\alphabeta(Y)}^{11} =
	\int\!\!\frac{d^4P}{(2\pi)^4}\int\!\!\frac{d^4K}{(2\pi)^4} e^{-i(P-K)\cdot(X-Y)} \mathcal{T}^\mnab(P,K)
	\label{corr.2}
\end{eqnarray}
where
\begin{eqnarray}
	\mathcal{T}^\mnab(P,K) = \Tr \Big[
	\Big( \mathcal{O}_+^\munu(P) - \mathcal{O}_-^\munu(-K) \Big) S^{11}(P) 
	\Big( \mathcal{O}_+^\alphabeta(K) - \mathcal{O}_-^\alphabeta(-P) \Big) S^{11}(K) \Big] \label{T.1}
\end{eqnarray}
in which the quantity $\mathcal{O}_\pm^\munu(P)$ is given by
\begin{eqnarray}
	\mathcal{O}_\pm^\munu(P) = \frac{1}{2}\TB{\gamma^\mu P^\nu + \gamma^\nu P^\mu -g^\munu (\cancel{P}\pm \mu_5\gamma^0\gamma^5 \mp M)}. \label{O.P}
\end{eqnarray}

We now substitute the correlation function from Eq.~\eqref{corr.2} into Eq.~\eqref{sp.fn} and obtain after performing some trivial integrations, the following expression of the in-medium spectral function in the static limit
\begin{eqnarray}
	\rho^\mnab = \lim\limits_{q^0\to0, \bm{q}=\bm{0}} \frac{\del}{\del q^0} \TB{ \tanh\FB{\frac{q^0}{2T}} \IM~i \int\!\!\frac{d^4K}{(2\pi)^4} \mathcal{T}^\mnab(P=Q+K,K)}. \label{sp.fn.2}
\end{eqnarray}
Substituting $\mathcal{T}^\mnab(P,K)$ from Eq.~\eqref{T.1} into Eq.~\eqref{sp.fn.2} followed by substituting the explicit expression of the propagator $S^{11}(P)$ and $S^{11}(K)$ from Eq.~\eqref{S11}, we obtain after evaluating the $dk^0$ integral and some standard algebra
\begin{eqnarray}
	\rho^\mnab &=& \lim\limits_{q^0\to0, \bm{q}=\bm{0}} \frac{\del}{\del q^0} \Big[ \tanh\FB{\frac{q^0}{2T}} (-\pi) \int\!\! \frac{d^3k}{(2\pi)^3} \sum_{r \in \{\pm\}} \sum_{s \in \{\pm\}} 
	\frac{1}{16rs\mu_5^2 pk} \frac{1}{4\wkr\wps} \nn \\ && 
	\times \Big[ \fb{1-\fkr -\fps +2\fkr\fps}\SB{\mathcal{N}_{P,K}^\mnab(k^0=-\wkr)\delta(q_0-\wkr-\wps) 
	+ \mathcal{N}_{P,K}^\mnab(k^0=\wkr) \delta(q_0+\wkr+\wps)} \nn \\ && 
	+ \fb{-\fkr -\fps +2\fkr\fps}\SB{\mathcal{N}_{P,K}^\mnab(k^0=\wkr)  \delta(q_0+\wkr-\wps) 
	+ \mathcal{N}_{P,K}^\mnab(k^0=-\wkr)  \delta(q_0-\wkr+\wps)} \Big]_{P=Q+K}\Big] \label{sp.fn.3}~,
\end{eqnarray}
 where, 
 \begin{eqnarray}
 	\mathcal{N}_{P,K}^\mnab = \Tr \Big[
 	\Big( \mathcal{O}_+^\munu(P) - \mathcal{O}_-^\munu(-K) \Big) \mathcal{D}(P) 
 	\Big( \mathcal{O}_+^\alphabeta(K) - \mathcal{O}_-^\alphabeta(-P) \Big) \mathcal{D}(K) \Big] \label{N.1}.
 \end{eqnarray}

In the expression of the spectral function $\rho^\mnab$ in Eq.~\eqref{sp.fn.3}, there exists sixteen Dirac delta functions in the integrand leading to the branch cuts of the spectral function in the complex energy $\angle q^0$-plane in the nonstatic limit. The terms with the first two delta functions inside square bracket of Eq.~\eqref{sp.fn.3} are called the unitary cuts whereas the terms with last two delta functions are termed as the Landau cuts. Each of the cuts further contains four subcuts corresponding to the different values of helicity tuple $(r,s)$. While calculating the viscous coefficients owing to a long-wavelength limit, the Unitary cuts as well as the Landau cuts with $r\ne s$ do not contribute so that Eq.~\eqref{sp.fn.3} finally becomes
\begin{eqnarray}
	\rho^\mnab &=& \lim\limits_{q^0\to0, \bm{q}=\bm{0}} \frac{\del}{\del q^0} \Bigg[ \tanh\FB{\frac{q^0}{2T}} \int\!\! \frac{d^3k}{(2\pi)^3} \sum_{r \in \{\pm\}}  
	\frac{1}{16\mu_5^2 pk} \frac{1}{4\wkr\wpr} \fb{\fkr +\fpr -2\fkr\fpr} \nn \\ &&  \hspace{-0.4cm}
	\times \lim\limits_{\Gamma_r\to 0} \Big[\mathcal{N}_{P,K}^\mnab(k^0=\wkr) 
		\frac{\Gamma_r}{(q_0+\wkr-\wpr)^2+\Gamma_r^2} 
		+ \mathcal{N}_{P,K}^\mnab(k^0=-\wkr) 
		 \frac{\Gamma_r}{(q_0-\wkr+\wpr)^2+\Gamma_r^2}
	\Big]_{P=Q+K} \Bigg] \label{sp.fn.4}
\end{eqnarray}
where we have introduced a Briet-Wigner representation of the Dirac delta function $\pi\delta(\omega) = \lim\limits_{\Gamma\to 0} \FB{\frac{\Gamma}{\omega^2+\Gamma^2}}$. The parameter $\Gamma_r$  appearing in Eq.~\eqref{sp.fn.4} can be thought as the spectral-width of the quark/antiquark of helicity $r$; and for a noninteracting system the spectral-width is zero leading to a Dirac delta  quark/antiquark spectral function. Moreover, in Eq.~\eqref{sp.fn.4}, while $\Gamma_r\to0$, the EMT spectral function $\rho^\mnab$ would diverge in the long-wavelength limit leading to a diverging viscous coefficients. This is the well known ``pinch singularity" problem~\cite{Hosoya:1983id} of the one-loop Green-Kubo formalism for a noninteracting system. This is due to the fact that the viscous coefficients are indeed infinite for a noninteracting system (like an ideal gas in which the mean free path diverges) and the momentum transport is effortless without any dissipation. Thus, in order to obtain a finite value for the transport coefficients, one must consider spectral broadening of quarks/antiquarks in thermal medium by means of taking finite values for their spectral-width $\Gamma_r>0$. We therefore identify $\Gamma_r = \Gamma_r(\bm{k};T,\mu_5)$ as the momentum, temperature and CCP dependent thermal width (which is equal to the inverse of the relaxation time $\tau(\bm{k};T,\mu_5)^{-1}$) of the quark/antiquark that enters into the expression of EMT spectral function in Eq.~\eqref{sp.fn.4} as a dynamical input so that Eq.~\eqref{sp.fn.4} simplifies to 
\begin{eqnarray}
	\rho^\mnab &=&  \frac{1}{2T} \int\!\! \frac{d^3k}{(2\pi)^3} \sum_{r \in \{\pm\}}  
	\frac{1}{32\mu_5^2 k^2(\wkr)^2}  \fkr\fb{1 -\fkr} \frac{1}{\Gamma_r(\bm{k};T,\mu_5)} 
	\TB{\mathcal{N}_{K,K}^\mnab(k^0=\wkr) + \mathcal{N}_{K,K}^\mnab(k^0=-\wkr)},  \label{sp.fn.5}
\end{eqnarray}
where the quantity $\mathcal{N}_{K,K}^\mnab$ is obtained in a simplified form from Eqs.~\eqref{N.1} and \eqref{O.P} as
\begin{eqnarray}
	\mathcal{N}_{K,K}^\mnab = \Tr \TB{ \SB{\gamma^\mu K^\nu + \gamma^\nu K^\mu - g^\munu (\cancel{K}+\mu_5\gamma^0\gamma^5 - M)} \mathcal{D}(K) 
	\SB{\gamma^\alpha K^\beta + \gamma^\beta K^\alpha - g^\alphabeta (\cancel{K}+\mu_5\gamma^0\gamma^5 - M)} \mathcal{D}(K)}. \label{N.2}
\end{eqnarray}
It can be shown from Eq.~\eqref{D} that $(\cancel{K}+\mu_5\gamma^0\gamma^5 - M) \mathcal{D}(K) = (K_+^2K_-^2-2M^2 K_+\cdot K_- + M^4) = \displaystyle{ \prod_{r \in \{\pm\}}^{}} \TB{(k^0)^2-(\wkr)^2}$ so that Eq.~\eqref{N.2} further simplifies to the following:
\begin{eqnarray}
	\mathcal{N}_{K,K}^\mnab(k^0=\pm\wkr) &=& \Tr \TB{ \frac{}{}\fb{\gamma^\mu k^\nu + \gamma^\nu k^\mu } \mathcal{D}(K) 
		\fb{\gamma^\alpha K^\beta + \gamma^\beta K^\alpha } \mathcal{D}(K)}_{k^0=\pm\wkr} \\
&=& 4N_cN_f \sum_{j \in \{\pm\}}^{}\Big[ \fb{K_{-j}^2-M^2}^2 
\FB{K_j^\mu K^\nu K_j^\alpha K^\beta + K_j^\nu K^\mu K_j^\alpha K^\beta + K_j^\mu K^\nu K_j^\beta K^\alpha + K_j^\nu K^\mu K_j^\beta K^\alpha} \nn \\
&& -M^2(K_j-K_{-j})^2 \FB{K_{-j}^\mu K^\nu K_j^\alpha K^\beta + K_{-j}^\nu K^\mu K_j^\alpha K^\beta + K_{-j}^\mu K^\nu K_j^\beta K^\alpha + K_{-j}^\nu K^\mu K_j^\beta K^\alpha}\Big]_{k^0=\pm\wkr}
\label{N.3}
\end{eqnarray}

Having obtained the EMT spectral function $\rho^\mnab$ in its static limit, the shear $\eta$ and bulk $\zeta$ viscous coefficients can now be easily obtained from the following relations~\cite{Huang:2011dc}:
\begin{eqnarray}
	\eta &=& \frac{1}{10} \FB{ \Delta_\mu^\sigma \Delta_\nu^\rho - \frac{1}{3}\Delta^{\sigma\rho}\Delta_\munu }
	\FB{ \Delta_{\sigma\alpha} \Delta_{\rho\beta} - \frac{1}{3}\Delta_{\sigma\rho}\Delta_\alphabeta } \rho^\mnab, \label{eta.1}\\
	\zeta &=& \FB{\frac{1}{3}\Delta_\munu+\theta ~U_\mu U_\nu}\FB{\frac{1}{3}\Delta_\alphabeta+\theta ~U_\alpha U_\beta}\rho^\mnab \label{zeta.1}
\end{eqnarray}
where $U^\mu$ is the four-velocity of the medium, $\Delta^\munu = (g^\munu-U^\mu U^\nu)$ is a projector orthogonal to $U$, and $\theta=c_s^2$ is the isentropic squared speed of sound in the medium. In the local rest frame (LRF) of the medium, $U^\mu_\text{LRF} \equiv (1,\bm{0})$. Substituting Eq.~\eqref{sp.fn.5} into Eqs.~\eqref{eta.1} and \eqref{zeta.1}, we finally arrive at the following expressions of shear and bulk viscosities at finite temperature and CCP:
\begin{eqnarray}
	\eta(T,\mu_5) &=&  \frac{2N_cN_f}{15T} \int\!\! \frac{d^3k}{(2\pi)^3} \sum_{r \in \{\pm\}}  
	\frac{1}{(\wkr)^2} \TB{k^2 (k+r\mu_5)^2}  \fkr\fb{1 -\fkr} \frac{1}{\Gamma_r(\bm{k};T,\mu_5)},  \label{eta}\\
	\zeta(T,\mu_5) &=& \frac{2N_cN_f}{9T} \int\!\! \frac{d^3k}{(2\pi)^3} \sum_{r \in \{\pm\}}  
	\frac{1}{(\wkr)^2}   \Big[
	k^4(1-3\theta)^2 + 2k^3r\mu_5(1-9\theta+18\theta^2) + k^2 \SB{ (1-18\theta+54\theta^2)\mu_5^2 - 6M^2\theta(1-3\theta)} \nn \\
	&& \hspace{4.5cm} - 6kr\mu_5\theta(1-6\theta)(\mu_5^2+M^2) + 9\theta^2(\mu_5^2+M^2)^2 \Big] \fkr\fb{1 -\fkr} \frac{1}{\Gamma_r(\bm{k};T,\mu_5)}. \label{zeta}
\end{eqnarray}
It now remains to calculate the thermal width $\Gamma_r(\bm{k};T,\mu_5)$ of the quark/antiquark and the isentropic squared speed of sound $\theta(T,\mu_5) = c_s^2$ at finite CCP which we will discuss in the next two subsections respectively.

In the limit $\mu_5\to0$, the above two expressions of $\eta$ and $\zeta$ reduce to 
\begin{eqnarray}
	\eta(T,\mu_5=0) &=&  \frac{4N_cN_f}{15T} \int\!\! \frac{d^3k}{(2\pi)^3}  
	\frac{k^4}{\wk^2} f_k\fb{1 -f_k} \frac{1}{\Gamma(\bm{k};T)},  \label{eta.0}\\
	\zeta(T,\mu_5=0) &=& \frac{4N_cN_f}{9T} \int\!\! \frac{d^3k}{(2\pi)^3}   
	\frac{1}{\wk^2}   \Big[
	k^2(1-3\theta) - 3\theta M^2 \Big]^2 f_k\fb{1 -f_k} \frac{1}{\Gamma(\bm{k};T)}. \label{zeta.0}
\end{eqnarray}
where $\wk=\sqrt{k^2+M^2}$, $f_k = f(\wk)$ and $\Gamma(\bm{k};T)$ is the thermal width of quark/antiquark in absence of CCP. It is to be noted that Eqs.~\eqref{eta.0} and \eqref{zeta.0} are the well  established expressions of shear and bulk viscosities calculated using the relaxation time approximation (RTA) of kinetic theory or Green-Kubo formalism at vanishing CCP~\cite{Chakraborty:2010fr}.


\section{Dynamical Input: Thermal Width}  \label{sec.width}
The thermal width of the quark/antiquark $\Gamma_s(\bm{p};T,\mu_5)$ enters into the expression of the viscous coefficients as a dynamical input. For calculating the width, we consider the $2\to2$ scattering processes among the quarks/antiquarks mediated via the collective mesonic modes in scalar ($\pi$) and pseudoscalar ($\sigma$) channels. The thermal width of quark/antiquark at finite CCP has already been calculated in detail in an earlier work~\cite{Ghosh:2024vje} and we will present some essential steps here for completeness. The expression of quark thermal width reads
\begin{eqnarray}
	\Gamma_s(\bm{p};T,\mu_5) = \tau_s(\bm{p};T,\mu_5)^{-1}= N_cN_f \int\!\! \frac{d^3k}{(2\pi)^3} \sum_{r \in \{\pm\}} \fkr\fb{1-\fkr} v^{rs}_\text{rel} \TB{\sigma_{qq\to qq}(\mans) + \sigma_{q\qbar\to q\qbar}(\mans)} \label{gam}
\end{eqnarray} 
where, $v^{rs}_\text{rel} = \MB{\frac{\bm{k}}{\wkr} - \frac{\bm{p}}{\wps}}$ is the relative velocity, $\sigma_{qq\to qq}(\mans) = \frac{1}{2}\cdot\frac{1}{64\pi^2\mans}\int d\Omega\overline{\mb{\scrM_{qq}}^2}$ and $\sigma_{q\qbar\to q\qbar}(\mans) = \frac{1}{64\pi^2\mans}\int d\Omega\overline{\mb{\scrM_{q\qbar}}^2}$ are the isospin averaged cross sections for the $2\to2$ scattering processes $q(K)q(P)\to q(K')q(P')$ and $q(K)\qbar(P)\to q(K')\qbar(P')$ respectively with $\overline{\mb{\scrM_{qq}}^2}$ and $\overline{\mb{\scrM_{q\qbar}}^2}$ being the corresponding spin-isospin averaged squared invariant amplitude, and $\mans=(\wkr+\wps)^2-(\bm{k}+\bm{p})^2$ is the Mandlestam variable. The explicit expressions of the squared matrix elements in MFA are as follows~\cite{Ghosh:2024vje}
\begin{eqnarray}
	\overline{\MB{\scrM_{qq}}^2} &=& 3\mant^2|D_\mant^\pi|^2 + 3\manu^2|D_\manu^\pi|^2 + (4M^2-\mant)^2|D_\mant^\sigma|^2 + (4M^2-\manu)^2|D_\manu^\sigma|^2 + \frac{1}{2N_C} \RE \big[ 3\mant\manu D_\mant^\pi D_\manu^{\pi*} \nn \\
	&& - 3\mant(4M^2-\manu) D_\mant^\pi D_\manu^{\sigma*} - 3\manu(4M^2-\mant) D_\mant^\sigma D_\manu^{\pi*} + \{(4M^2-\mant)(4M^2-\manu)+2\mant\manu\} D_\mant^\sigma D_\manu^{\sigma*}  \big], \label{M2qq}\\
	\overline{\MB{\scrM_{q\qbar}}^2} &=& 3\mans^2|D_\mans^\pi|^2 + 3\mant^2|D_\mant^\pi|^2 + (4M^2-\mant)^2|D_\mant^\sigma|^2 + (4M^2-\mans)^2|D_\mans^\sigma|^2 + \frac{1}{2N_C} \RE \big[ 3\mans\mant D_\mans^\pi D_\mant^{\pi*} \nn \\
	&& - 3\mans(4M^2-\mant) D_\mans^\pi D_\mant^{\sigma*}  - 3\mant(4M^2-\mans) D_\mans^\sigma D_\mant^{\pi*} + \{(4M^2-\mans)(4M^2-\mant)-2\mans\mant\} D_\mans^\sigma D_\mant^{\sigma*}  \big] \label{M2qqb}~,
\end{eqnarray}
where $\mant=(K-K')^2$, $\manu=(K-P')^2$ are the Mandlestam variables, $D_{\mans}^h = D^h(\rs,0;T,\mu_5)$, $D_{\mant}^h = D^h(0,\sqrt{-\mant};T,\mu_5)$ and $D_{\manu}^h = D^h(0,\sqrt{-\manu};T,\mu_5)$ with $h\in \{ \pi,\sigma \}$ and $D^h(q^0,q;T,\mu_5)$ being the in-medium complete mesonic propagators given by
\begin{eqnarray}
		D^h(q^0,q;T,\mu_5) = \frac{2G}{1-2G\RE\Pi^{11}_h(q^0,q;T,\mu_5)-2iG\text{sign}(q^0)\tanh\FB{\frac{q^0}{2T}}\IM\Pi^{11}_h(q^0,q;T,\mu_5)} \label{Dh.f}~,
\end{eqnarray}
in which $\Pi^{11}_h(q^0,q;T,\mu_5)$ is $11$-component of the real time thermal polarization function at finite CCP in the mesonic channel $h\in \{ \pi,\sigma \}$. They are given by
\begin{eqnarray}
	\Pi_h^{11}(q^0,q;T,\mu_5) = i\int\!\!\frac{d^4K}{(2\pi)^4}\Tr\TB{\Gamma_hS_{11}(P=Q+K)\Gamma_hS_{11}(K)} \label{Pi.11}~,
\end{eqnarray}
where $\Gamma_\pi = i\gamma^5$, $\Gamma_\sigma = \identity$ and the quark propagator $S_{11}(K)$ is defined in Eq.~\eqref{S11}. On substituting Eq.~\eqref{S11} into Eq.~\eqref{Pi.11} and performing the $dk^0$ integration, we obtain after some standard algebra the following expressions for the real and imaginary parts of the 11-component of mesonic polarization functions at finite CCP:
\begin{eqnarray}
\RE\Pi^{11}_h(q^0,0;T,\mu_5) &=& \int_{0}^{\infty}\!\! \frac{dk}{2\pi^2} \sum_{r \in \{\pm\}} \sum_{s \in \{\pm\}} 
	\frac{1}{64rs\mu_5^2} 
	\mathcal{P}\Big[ \frac{\mathcal{N}_h(k^0=\wkr)(f_\Lambda(k)-2\fkr)}{\wkr \SB{(q^0+\wkr)^2-(\wks)^2}}	
	+ \frac{\mathcal{N}_h(k^0=-\wkr)(f_\Lambda(k)-2\fkr)}{\wkr \SB{(q^0-\wkr)^2-(\wks)^2}} \nn \\	
	&& + \frac{\mathcal{N}_h(k^0=-q^0+\wks)(f_\Lambda(k)-2\fks)}{\wks \SB{(q^0-\wks)^2-(\wkr)^2}}
	+ \frac{\mathcal{N}_h(k^0=-q^0-\wks)(f_\Lambda(k)-2\fks)}{\wks \SB{(q^0+\wks)^2-(\wkr)^2}} \Big]_{\bm{q}=\bm{0}}, \\
	\label{repi.1}
	\IM\Pi^{11}_h(q^0,0;T,\mu_5) &=& - \int_{0}^{\infty}\!\! \frac{dk}{2\pi} \sum_{r \in \{\pm\}} \sum_{s \in \{\pm\}} 
	\frac{1}{16rs\mu_5^2} \frac{1}{4\wkr\wks} \nn \\ && 
	\times \Big[ \fb{1-\fkr -\fks +2\fkr\fks} \SB{ \mathcal{N}_h(k^0=-\wkr) \delta(q_0-\wkr-\wks) + \mathcal{N}_h(k^0=\wkr)   \delta(q_0+\wkr+\wks) } \nn \\ && 
	+ \fb{-\fkr -\fks +2\fkr\fks} \SB{\mathcal{N}_h(k^0=\wkr) \delta(q_0+\wkr-\wks) + \mathcal{N}_h(k^0=-\wkr)  \delta(q_0-\wkr+\wks)} \Big]_{\bm{q}=\bm{0}}, \\
	\label{impi.1}
	\RE\Pi^{11}_h(0,q;T,\mu_5) &=& \int_{0}^{\infty} \!\! \frac{dk}{2\pi^2} k \sum_{r \in \{\pm\}} \sum_{s \in \{\pm\}} 
	\frac{\fb{f_\Lambda(k)-2\fkr}}{32rs\mu_5^2\wkr}  \int_{-1}^{1}d(\cos\theta^{\bm{q}}_{\bm{k}}) \frac{1}{p}
	\mathcal{P}\Big[ \frac{\mathcal{N}_h(k^0=\wkr)}{(\wkr)^2-(\wps)^2} \Big]_{\bm{p}=\bm{q}+\bm{k}, q^0=0}, \label{repi.2} \\
	\IM\Pi^{11}_h(0,q;T,\mu_5) &=& 0, \label{impi.2}
\end{eqnarray}
where $\theta^{\bm{q}}_{\bm{k}}$ is the angle between $\bm{q}$ and $\bm{k}$, and the quantity $\mathcal{N}_h(K,Q)$ is
\begin{eqnarray}
	\mathcal{N}_h(K,Q) &=& N_cN_f\Big[4a_hM^6 + 2M^4 \big\{-2a_h (K_+\cdot K_-)+(K_+\cdot P_-)+(K_-\cdot P_+)-2a_h(P_+\cdot P_-)\big\} \nn \\
	&& -2M^2 \big\{ -2a_h(K_+\cdot P_-)(K_+\cdot P_+) -2a_h(K_+\cdot K_-)(P_+\cdot P_-) + 2a_h(K_+\cdot P_+)(K_-\cdot P_-) \nn \\
	&&+ (K_-\cdot P_-)(K_+^2+P_+^2) + (K_+\cdot P_+)(K_-^2+P_-^2) \big\} + 2 \big\{ (K_+\cdot P_-)K_-^2P_+^2 + (K_-\cdot P_+)K_+^2P_-^2 \big\}\Big]_{P=Q+K} \label{Nh}~,
\end{eqnarray} 
in which $a_\pi=-1$ and $a_\sigma=1$. In Eq.~\eqref{repi.1} and \eqref{repi.2}, $\mathcal{P}$ denotes the Cauchy principal value integration. 


\section{Thermodynamic Quantities and Speed of Sound}  \label{sec.thermodynamics}
In order to calculate the bulk viscosity $\zeta$ from Eq.~\eqref{zeta}, we also need the quantity $\theta=c_s^2 = \FB{\dfrac{\del P}{\del\varepsilon}}_s$ which is the isentropic squared speed of sound in the medium where $P$, $\varepsilon$ and $s$ are respectively the physical pressure, physical energy density and physical entropy density of the thermodynamic system in consideration. In NJL model, such quantities can be obtained from the thermodynamic potential density $\Omega$ which is given by
\begin{eqnarray}
	\Omega(T,\mu_5) = \frac{(M-m)^2}{4G} - 2N_cN_f \int\!\!\frac{d^3p}{(2\pi)^3} \sum_{r \in \{\pm\}} \TB{\frac{1}{2}\wpr f_\Lambda(p) + T\ln\FB{1+e^{-\wpr/T}}}.
\end{eqnarray}
The physical potential density $\widetilde{\Omega}(T,\mu_5)$ is measurable only with respect to the vacuum $\Omega_\text{vac} = \Omega(T=0,\mu_5)$ and is given by~\cite{Zhuang:1994dw}
\begin{eqnarray}
	\widetilde{\Omega}(T,\mu_5)  &=& \Omega(T,\mu_5) - \Omega(T=0,\mu_5) \nn \\
	&=& \frac{(M-m)^2}{4G} - \frac{(M_0-m)^2}{4G} - 2N_cN_f \int\!\!\frac{d^3p}{(2\pi)^3} \sum_{r \in \{\pm\}} \TB{\frac{1}{2}(\wpr-\wopr)f_\Lambda(p) + T\ln\FB{1+e^{-\wpr/T}}},
\end{eqnarray}
where $M_0 = M(T=0,\mu_5)$ is the constituent quark mass at $T=0$ and $\wopr = \sqrt{(p+r\mu_5)^2+M_0^2}$. 
Having obtained the physical potential density $\widetilde{\Omega}(T,\mu_5)$, all the physical thermodynamic quantities can be calculated as follows. The pressure is given by $P(T,\mu_5) = -\widetilde{\Omega}$ and the entropy density is given by
\begin{eqnarray}
	s(T,\mu_5) &=& -\FB{\frac{\del\widetilde{\Omega}}{\del T}} = 2N_cN_f \int\!\!\frac{d^3p}{(2\pi)^3} \sum_{r \in \{\pm\}} \TB{ \frac{\wpr}{T}\fpr + \ln\FB{1+e^{-\wpr/T}}}. \label{s}
\end{eqnarray}
From $P(T,\mu_5)$ and $s(T,\mu_5)$, we can calculate the energy density from $\varepsilon(T,\mu_5) = (Ts-P)$, from which we obtain the isentropic squared speed of sound as 
$\theta=c_s^2 = \FB{\frac{\del P}{\del\varepsilon}}_s = \FB{\frac{\del P}{\del T}} \Big/ \FB{{\frac{\del \varepsilon}{\del T}}} = \frac{s}{C_V}$
where $C_V = \FB{{\frac{\del \varepsilon}{\del T}}}$ is the specific heat given by
\begin{eqnarray}
	C_V &=& 2N_cN_f \int\!\!\frac{d^3p}{(2\pi)^3} \sum_{r \in \{\pm\}} \fpr(1-\fpr) \TB{\frac{M}{T}\frac{\del M}{\del T}- \FB{\frac{\wpr}{T}}^2}, ~~\text{in which,}\\
	\FB{{\frac{\del M}{\del T}}} &=& \frac{2G}{1-2G\mathcal{A}} 2N_cN_f \int\!\!\frac{d^3p}{(2\pi)^3} \sum_{r \in \{\pm\}} \frac{M}{T^2} \fpr(1-\fpr), \\
	\mathcal{A} &=& 2N_cN_f \int\!\!\frac{d^3p}{(2\pi)^3} \sum_{r \in \{\pm\}} \frac{1}{\wpr} 
	\TB{ \FB{\frac{1}{2}f_\Lambda(p)-\fpr}\FB{1-\frac{M^2}{(\wpr)^2}} - \frac{M^2}{T\wpr}\fpr(1-\fpr) }.
\end{eqnarray}

\section{Numerical Results And Discussions} \label{sec.results}
Before discussing the numerical results, let us first mention the values of the different parameters of the NJL model that have been used in this work. The parameters are chosen to reproduces the experimental/phenomenological vacuum ($T=\mu_5=0$) values of the quark-condensate per flavor $\ensembleaverage{\psibar\psi}/N_f$, pion decay constant $f_\pi$, constituent quark mass in vacuum $M(T=\mu_5=0)$ and the pion mass $m_\pi$. The values of the parameters as well the experimental/phenomenological vacuum values of different quantities are tabulated in Table-\ref{tab.1}
\begin{table}[h]
	\begin{tabular}[c]{c c c}
		$m$ (MeV) & $G$ (GeV$^2$)& $\Lambda$ (MeV) \\
		\hline \hline
		~~~~~~5.6~~~~~~& ~~~~~~5.742~~~~~~ & ~~~~~~568.69~~~~~~ \\
		\hline
	\end{tabular} \hspace{1.5cm}
	\begin{tabular}[c]{cccc}
	$\ensembleaverage{\psibar\psi}/N_f$ (MeV$^3$) & $f_\pi$ (MeV)& ~~~$M(T=\mu_5=0)$ (MeV)~~~ & $m_\pi$ (MeV) \\
	\hline \hline
	~~~~~~~~~~$-(245.1)^3$~~~~~~~~~~ & ~~~~~~~~91.9~~~~~~~~ & 343.6 & ~~~~~~~~140~~~~~~~~ \\
	\hline
\end{tabular}
\caption{Table showing the values to the parameters of the NJL model used in this work and the respective experimental/phenomenological vacuum values of different quantities.}
\label{tab.1}
\end{table}
\begin{figure}[h]
	\includegraphics[angle=0,scale=0.35]{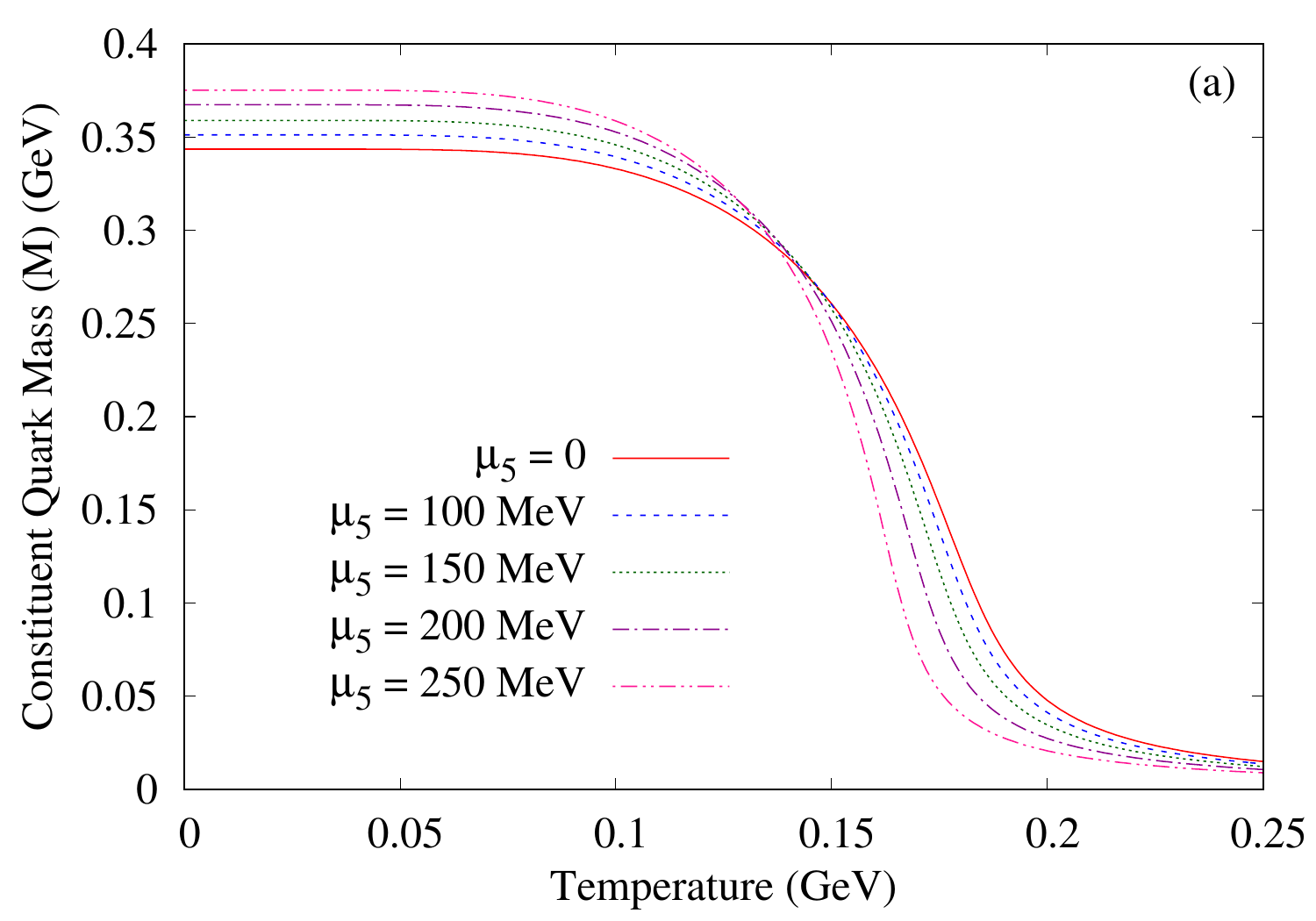} \includegraphics[angle=-0,scale=0.35]{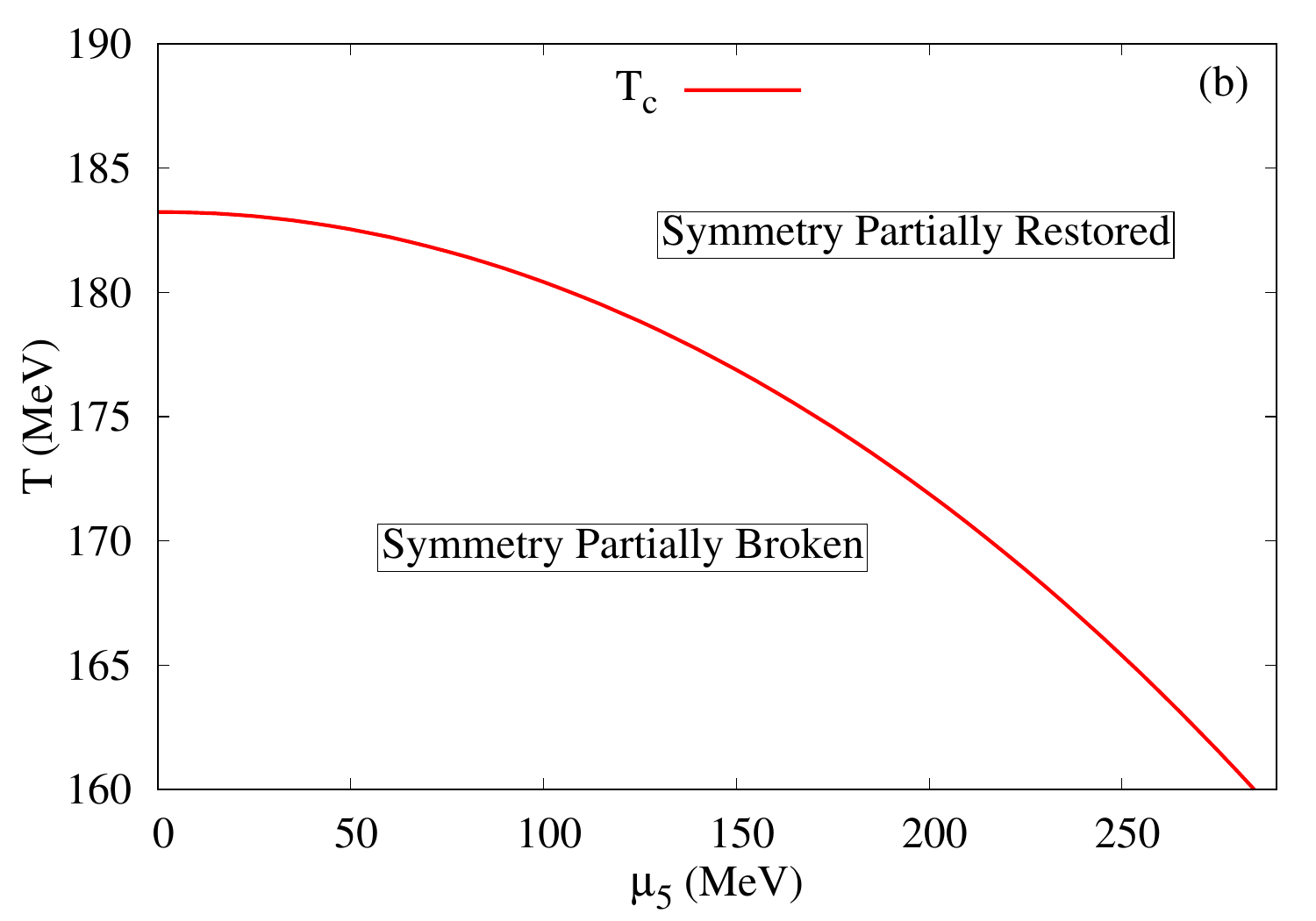}
	\caption{(Color Online) The variation of the (a) constituent quark mass $M$ as a function of temperature at different $\mu_5$, and (b) transition temperature $T_c$ as a function of $\mu_5$ yielding the chiral phase diagram on $T-\mu_5$ plane.}
	\label{fig.M}
\end{figure}

We now depict in Fig.~\ref{fig.M}(a) the variation of the constituent quark mass $M=M(T,\mu_5)$ as a function of temperature at different values of $\mu_5$. We notice that $M$ is large at the low temperature region owing to a stronger quark condensate leading to spontaneously broken chiral symmetry. With the increase in temperature, $M$ remains almost constant up to a certain $T$, followed by a sudden decrease around the pseudochiral phase transition and then $M$ asymptotically approaches the current quark mass value $m$ at the high temperature region as a consequence of the partial restoration of the chiral symmetry. On the other hand, with the increase in CCP, $M$ shows distinct behavior in the low and high temperature region. In particular, $M$ increases (decreases) with the increase in $\mu_5$ in the low (high) $T$-region making the chiral condensate stronger (weaker). Moreover, at finite CCP, the sudden drop in $M$ vis-a-vis partial restoration of the chiral symmetry occurs at a lower temperature as compared to the zero CCP case leading to the fact that a chiral imbalance blocks/hinders the dynamical symmetry breaking known as the inverse chiral catalysis (ICC) effect~\cite{Gatto:2011wc,Chernodub:2011fr,Ruggieri:2011xc,Yu:2015hym,Fukushima:2010fe,Ghosh:2022xbf,Chaudhuri:2022rwo,Ghosh:2023rft}. Next in Fig.~\ref{fig.M}(b), we have shown the chiral phase diagram on the $T-\mu_5$ plane along with the variation of the pseudochiral phase transition temperature ($T_c$) as a function of CCP. The transition temperature has been calculated from the position of the peak of the chiral susceptibility $\chi = \frac{1}{2G}\FB{\frac{\del M}{\del m}-1}$. As can be observed in the figure, $T_c$ decreases monotonically with the increase in CCP owing to the ICC effect. This is in line with most of the conventional NJL type calculations~\cite{Gatto:2011wc,Chernodub:2011fr,Ruggieri:2011xc,Yu:2015hym,Fukushima:2010fe} in sharp contrast to LQCD simulations~\cite{Braguta:2015zta,Braguta:2015owi} which show the opposite trend, i.e. the transition temperature increases with CCP indicating a chiral catalysis (CC). It is to be noted that the conventional 3-momentum cut-off regularization scheme produces a CC effect as a regularization artifact~\cite {Fukushima:2010fe}. Again, Ref.~\cite{Andrianov:2013dta}, using a two-flavor NJL model with broken $U_A(1)$ symmetry shows a CC effect which is also obtained in Ref.~\cite{Ruggieri:2016ejz}  using a non-local  NJL model with different interaction kernels. Recently, the Lattice data for the chiral and deconfinement critical temperatures as a function of the CCP have been reproduced by including a $\mu_5$-dependence in the parametrization of the quadratic $\Phi^\dagger \Phi$ term in the Polyakov-loop potential along with the usual temperature dependence~\cite{Azeredo:2024sqc}. However this comparison  should be taken with caution because the  LQCD data has been obtained with a large pion mass without extrapolating to the physical limit.
\begin{figure}[h]
	\begin{center}
	\includegraphics[angle=-0,scale=0.7]{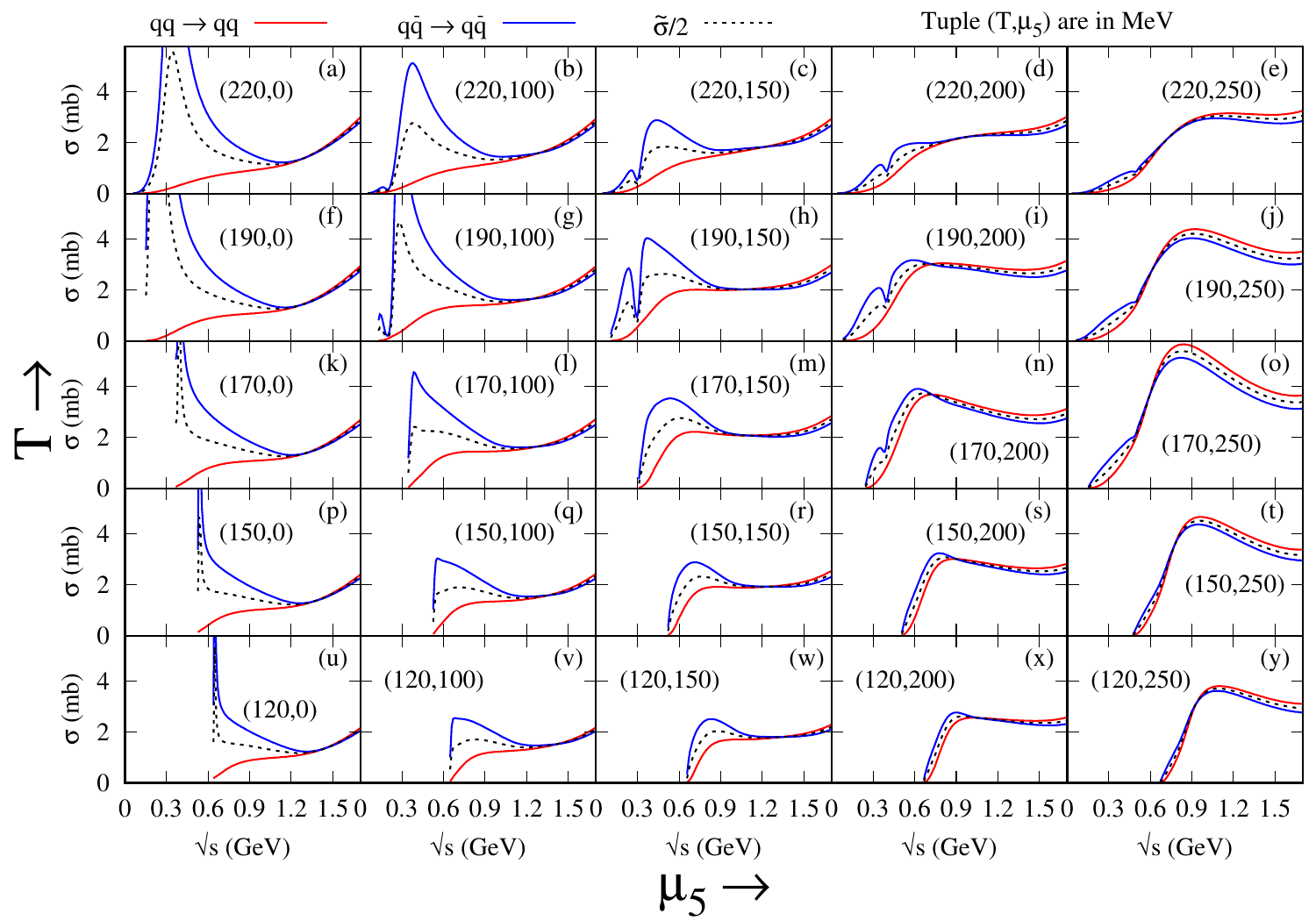} 
	\caption{(Color Online) (a)-(y) The variation of the isospin averaged $qq\to qq$ and $q\qbar\to q\qbar$ cross sections as a function of center of mass energy $\rs$ projected on a discretized $T-\mu_5$ plane containing different combinations of the temperature-CCP tuple ($T,\mu_5$). The CCP (temperature) increases along the horizontally rightward (vertically upward) direction. The total of cross sections $ \widetilde{\sigma} = (\sigma_{qq\to qq}+\sigma_{q\qbar\to q\qbar})$ scaled by a factor $\frac{1}{2}$ is also shown by black-dashed line.}
	\label{fig.xsection}
	\end{center}
\end{figure}

We now turn our attention to the numerical results of the $2\to2$ elastic cross sections among the quarks/antiquarks. For this we have shown in Figs.~\ref{fig.xsection}(a)-(y), the variation of the isospin-averaged total $qq\to qq$ and $q\qbar\to q\qbar$ cross sections along with their total as a function of center of mass energy $\rs$. The cross sections are calculated for different combinations of the temperature-CCP tuple ($T,\mu_5$) and the graphs have been arranged on a discretized $T-\mu_5$ plane capturing the different stages of pseudo-chiral phase transition. In those panels/subfigures of Fig.~\ref{fig.xsection}, the CCP (temperature) increases along the horizontally rightward (vertically upward) direction as indicated by large font texts at the bottom and left margin of the figure. 

In Fig.~\ref{fig.xsection}, we first notice a temperature and CCP dependent threshold of the center of mass energy $\rs_\text{min}=2M(T,\mu_5)$. The threshold $\rs_\text{min}$  decreases significantly with the increase in temperature as we move vertically upward from lower to higher panels. The CCP dependence of $\rs_\text{min}$ is weaker as compared to the $T$-dependence as can be seen if we move horizontally rightward from left to right panels. These behavior is easily understandable from the $T$ and $\mu_5$ dependence of $M(T,\mu_5)$ in Fig.~\ref{fig.M}(a). 

We also notice a significantly distinct $\rs$-dependence between $\sigma_{q\qbar\to q\qbar}$ and $\sigma_{qq\to qq}$ mainly in the low CCP regions; in particular $\sigma_{q\qbar\to q\qbar}$ shows a Breit-Wigner like structure in its $\rs$ dependence whereas $\sigma_{qq\to qq}$ shows no such structure. This is due to the exchange of mesonic resonance in the $\mans$-channel diagrams of the process $\sigma_{q\qbar\to q\qbar}$. With the increase in temperature, the exchanged of mesonic mode acquires significant width owing to their spectral broadening in thermal medium~\cite{Mallik:2016anp}. This in turn leads to the increase in the widths of the Breit-Wigner structures of $\sigma_{q\qbar\to q\qbar}(\rs)$ with the increase in temperature. Next we observe that, with the increase in CCP, the Breit-Wigner structures of $\sigma_{q\qbar\to q\qbar}(\rs)$ is destroyed. This is mainly due to the Landau-damping of the timelike mesonic resonances, a purely finite CCP effect~\cite{Ghosh:2023rft}, for which the mesons develop significant thermal-widths beyond their regular pole positions; the same is also responsible for the double peak structures in $\sigma_{q\qbar\to q\qbar}(\rs)$ observed visually in some of the panels at finite CCP.

We further observe in Fig.~\ref{fig.xsection} that, $\sigma_{qq\to qq}$ monotonically increases with the increase in $\rs$ in low CCP region; whereas it becomes nonmonotonic at higher values of CCP. Moreover  $\sigma_{qq\to qq}$ has a weak temperature dependence in all the CCP-region as can be noticed by comparing panels along vertical direction. $\sigma_{qq\to qq}$ is found to be maximum near the chiral phase transition region as seen in the middle horizontal panel and the overall magnitude of $\sigma_{qq\to qq}$ increases with the increase in CCP for all the temperatures. Away from the peak positions of the Breit-Wigner structures of $\sigma_{q\qbar\to q\qbar}$, the temperature and CCP dependence of $\sigma_{q\qbar\to q\qbar}$ is found to be similar to that of $\sigma_{qq\to qq}$. 

We have also shown the total cross section $\widetilde{\sigma} =  \sum_{x }^{} \sigma_{qx\to qx} = (\sigma_{qq\to qq}+\sigma_{q\qbar\to q\qbar}) $ scaled by a factor $\frac{1}{2}$ by black-dashed line as it goes directly into the expression of quark relaxation time in Eq.~\eqref{gam}.  At smaller values of $\rs$ near the threshold, $\widetilde{\sigma}$ is dominated by $\sigma_{q\qbar\to q\qbar}$ owing to its Briet-Wigner peak in the low CCP-region. Thus, we notice that $\widetilde{\sigma}$ has an overall significant (insignificant) temperature dependence in the low (high) CCP-region.
\begin{figure}[h]
	\begin{center}
		\includegraphics[angle=-0,scale=0.7]{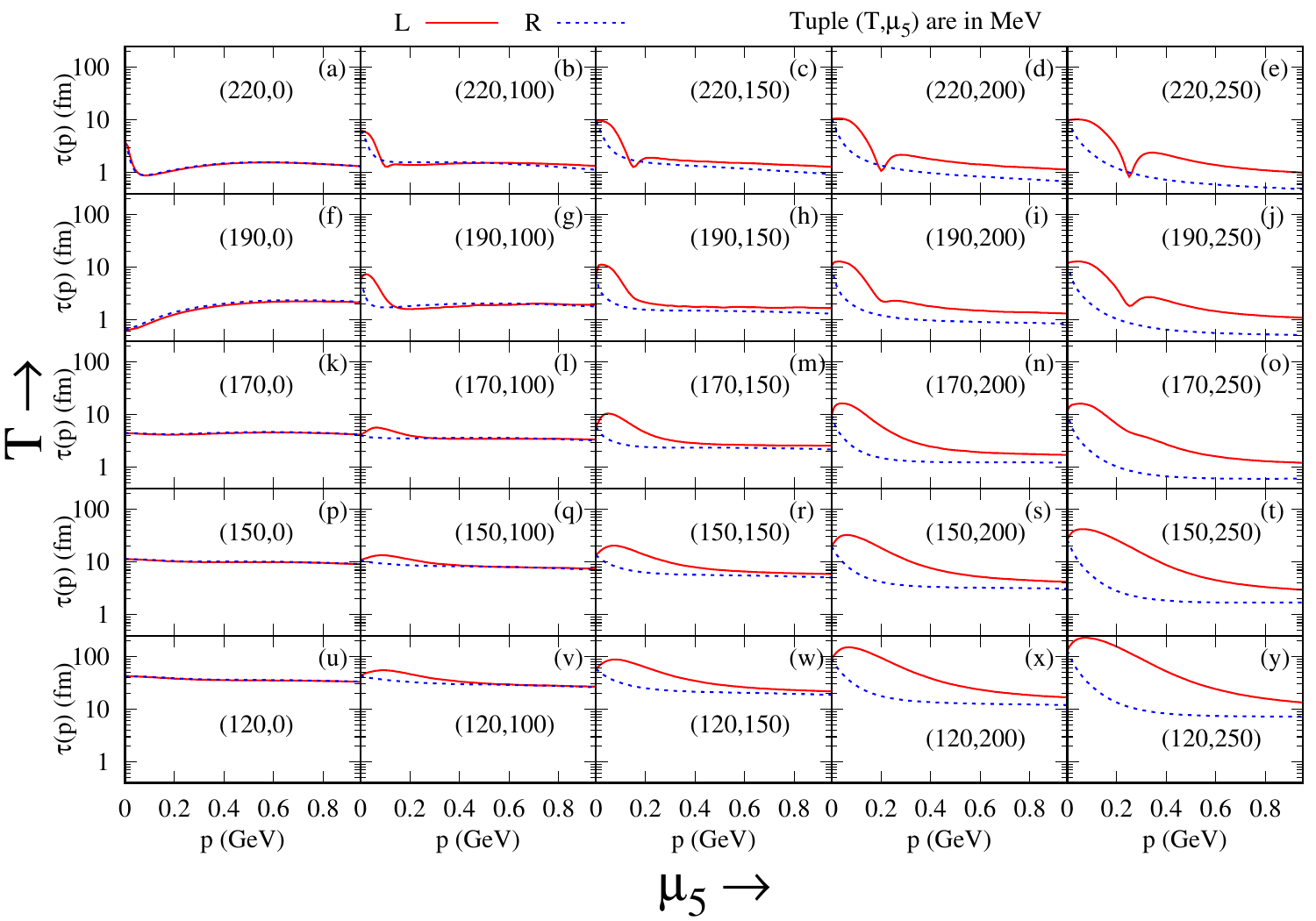} 
		\caption{(Color Online) (a)-(y) The variation of the quark relaxation time $\tau(p;T,\mu_5)$ as a function of momentum $p$ projected on a discretized $T-\mu_5$ plane containing different combinations of the temperature-CCP tuple ($T,\mu_5$). The CCP (temperature) increases along the horizontally rightward (vertically upward) direction.}
		\label{fig.tau}
	\end{center}
\end{figure}

Having discussed the cross section, we now move to the numerical results of the relaxation time $\tau_s(p;T,\mu_5)$ which has been obtained from Eq.~\eqref{gam}. In Figs.~\ref{fig.tau}(a)-(y), we have depicted the variation of the quark relaxation times $\tau_- \equiv \tau_L(p;T,\mu_5)$ and $\tau_+ \equiv \tau_R(p;T,\mu_5)$ as a function of momentum $p$ for different combinations of the temperature-CCP tuple ($T,\mu_5$) projected on a discretized $T-\mu_5$ plane similar to Fig.~\ref{fig.xsection}. To explain the behavior of $\tau$, we first note from Eq.~\eqref{gam}, that $\tau_s \sim \frac{1}{n\widetilde{\sigma}}$ where $n$ is the quark number density. Thus the momentum dependence of $\tau_s(p)$ mainly comes from the momentum/energy dependence of $\widetilde{\sigma}(\mans)$ with $\mans \sim (\wps)^2 = (p+ s\mu_5)^2+M^2$; some $p$-dependence also comes from the relative velocity term of Eq.~\eqref{gam}. We also note that at vanishing CCP, $\tau_L = \tau_R$.

Now, at vanishing CCP, $\rs$ increases monotonically with the increase in $p$. Observing Figs.~\ref{fig.xsection}(k), (p) and (u), we see that, except for an initial narrow spike, $\widetilde{\sigma}$ is almost independent of $\rs \sim p$ throughout the whole range of $\rs$ or thus independent of $p$. This explains the fact that $\tau_s(p)$ is almost independent of $p$ in Figs.~\ref{fig.tau}(k), (p) and (u). On the contrary, in Fig.~\ref{fig.tau}(a) we notice a minimum in $\tau_s$ at low $p$-region followed by an almost $p$-independent nature in high $p$-region. The same also will show up in Fig.~\ref{fig.tau}(f) when the low $p$ region is enlarged. This can be explained by looking at Figs.~\ref{fig.xsection}(f) and (a), where at lower $\rs \sim p$,  $\widetilde{\sigma}$ has a maximum (the peak of the Breit-Wigner structure) which translates into a minimum in $\tau_s$ in the low $p$-region; whereas in the higher $\rs \sim p$-region, $\widetilde{\sigma}$ is almost independent of $\rs \sim p$ yielding an almost $p$-independent $\tau_s$ for high $p$-region. At finite CCP $\tau_s$ has an overall decreasing trend with the increase in $p$ which is due to the fact that $\widetilde{\sigma}$ shows an overall increasing trend with the increase in $\rs$ as well as due to the $p$-dependence of the relative velocity term of Eq.~\eqref{gam}. 

Having discussed the momentum dependence of $\tau_s$, we now move to the temperature and CCP dependence of $\tau_s \sim \frac{1}{n\widetilde{\sigma}} $ which mainly comes from the temperature and CCP dependence of the number density $n$ of the quarks/antiquarks. In the massless case, we have 
\begin{eqnarray}
	n(T,\mu_5; M=0) &=& 2N_c N_f\sum_{r\in \{\pm\} }^{} \int \!\! \frac{d^3k}{(2\pi)^3} \frac{1}{e^{|k+r\mu_5|/T}+1} = \frac{N_c N_f}{\pi^2} T^3 \TB{ 3\zeta(3) +\ln(4)\FB{\frac{\mu_5}{T}}^2 } 
\end{eqnarray}
where $\zeta(x)$ is the Riemann zeta function. Thus, at leading order, $n \sim T^3\SB{a+b \FB{\frac{\mu_5}{T}}^2 }$ where $a$ and $b$ are some constants. 
This in turn makes $\tau_s \sim \frac{1}{n\widetilde{\sigma}} \sim \frac{1}{T^3\SB{a+b \FB{\frac{\mu_5}{T}}^2 }}$. 
Therefore, with the increase in temperature and CCP, the overall magnitude of $\tau_s$ decreases in most of the $p$-regions as can clearly be observed in Fig.~\ref{fig.tau}. However, in the low $p$-region (near the peak position), $\tau_s$ increases with the increase in CCP which may be attributed to the finite mass and additional CCP-dependence of $\widetilde{\sigma}$ and the relative velocity. 
\begin{figure}[h]
	\begin{center}
		\includegraphics[angle=-0,scale=0.35]{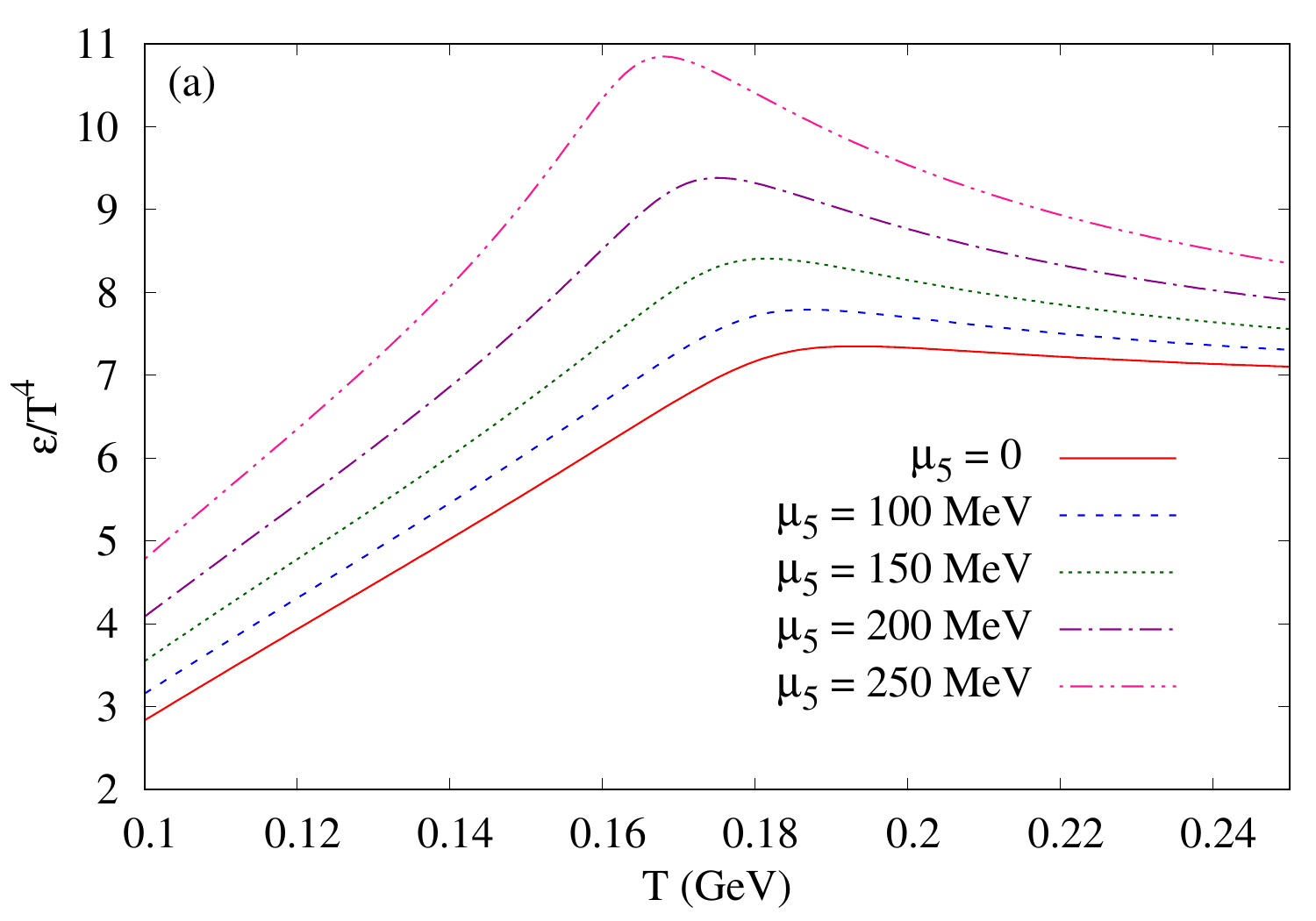} \includegraphics[angle=-0,scale=0.35]{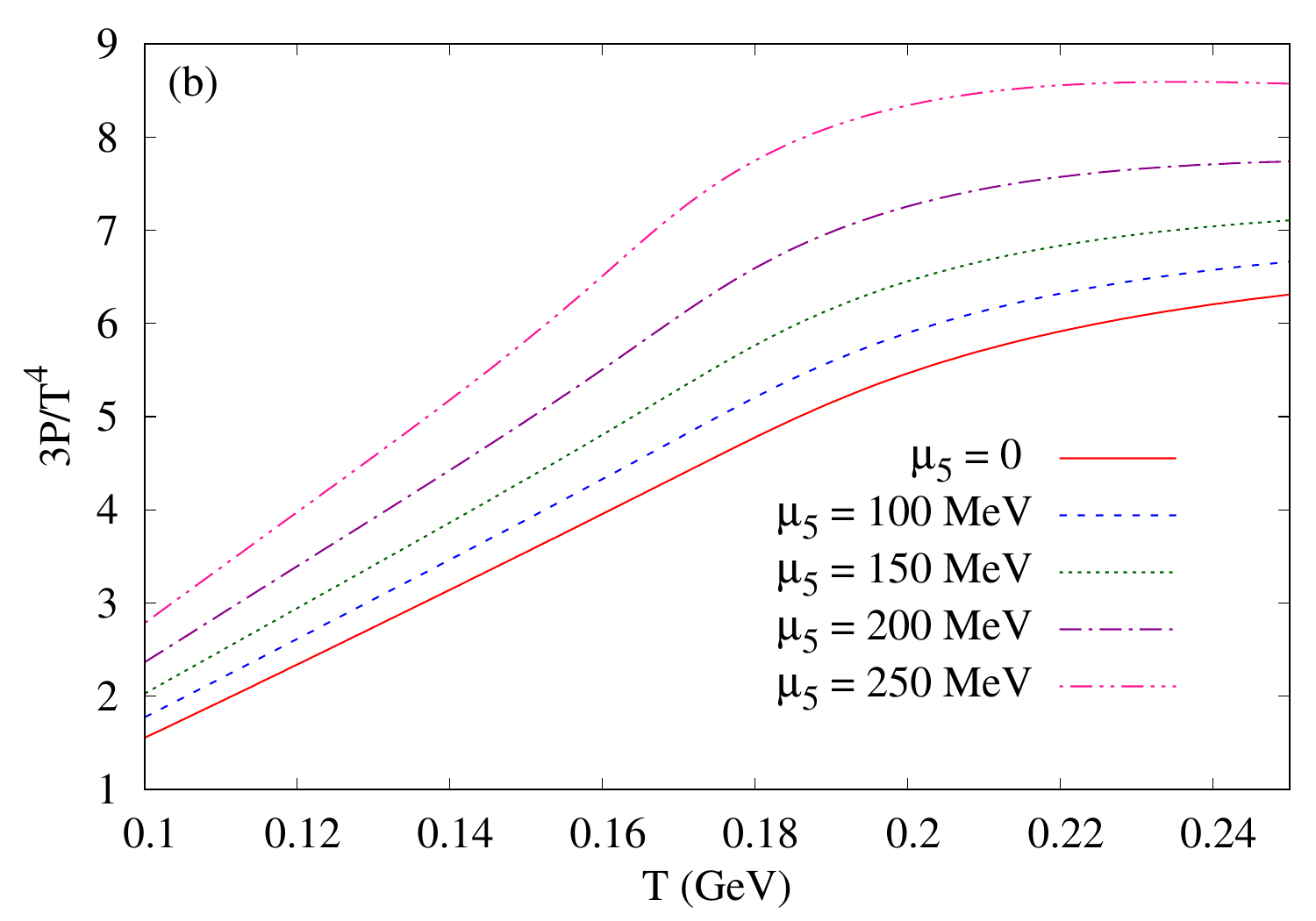}
		\includegraphics[angle=-0,scale=0.35]{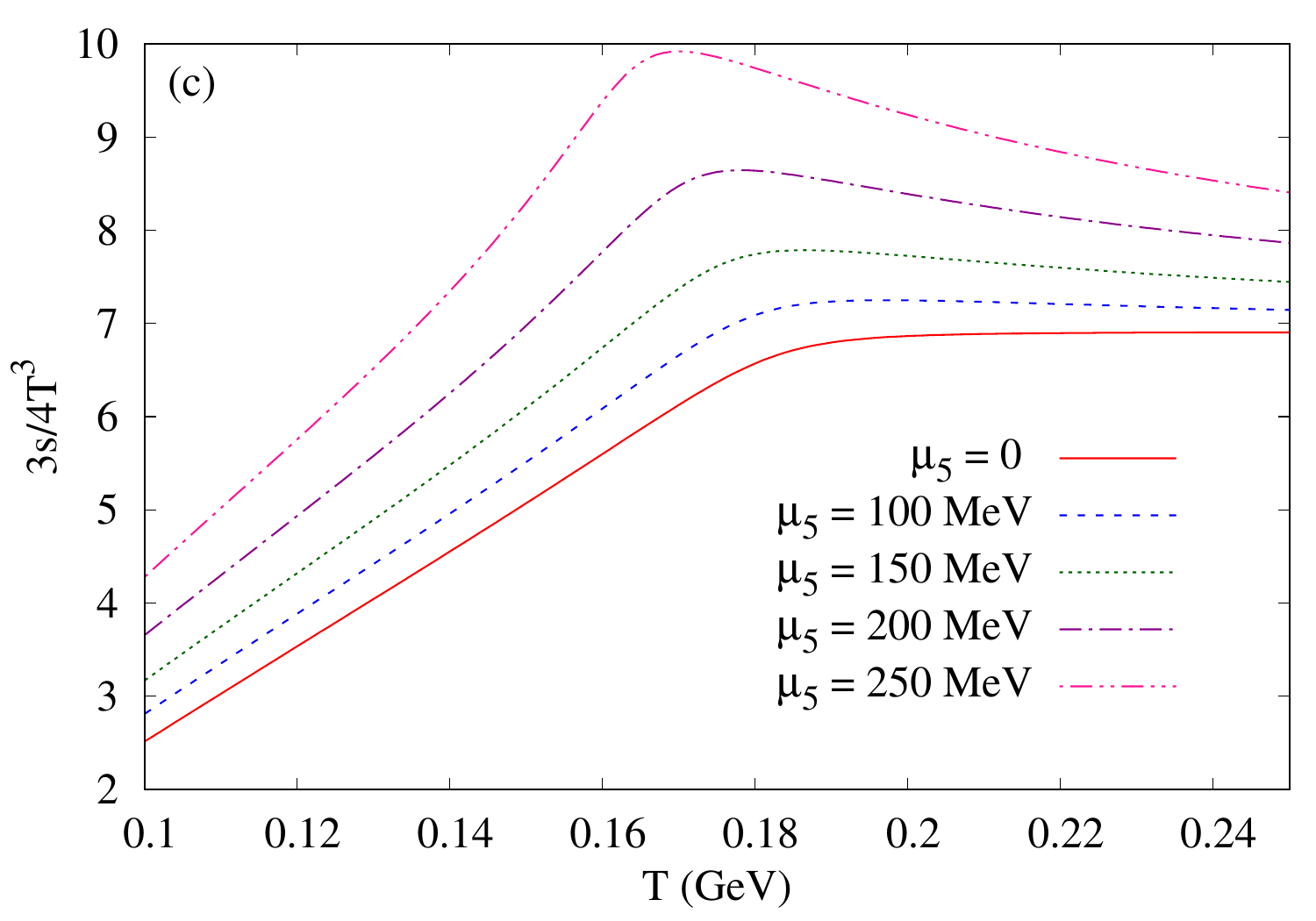}
		\includegraphics[angle=-0,scale=0.35]{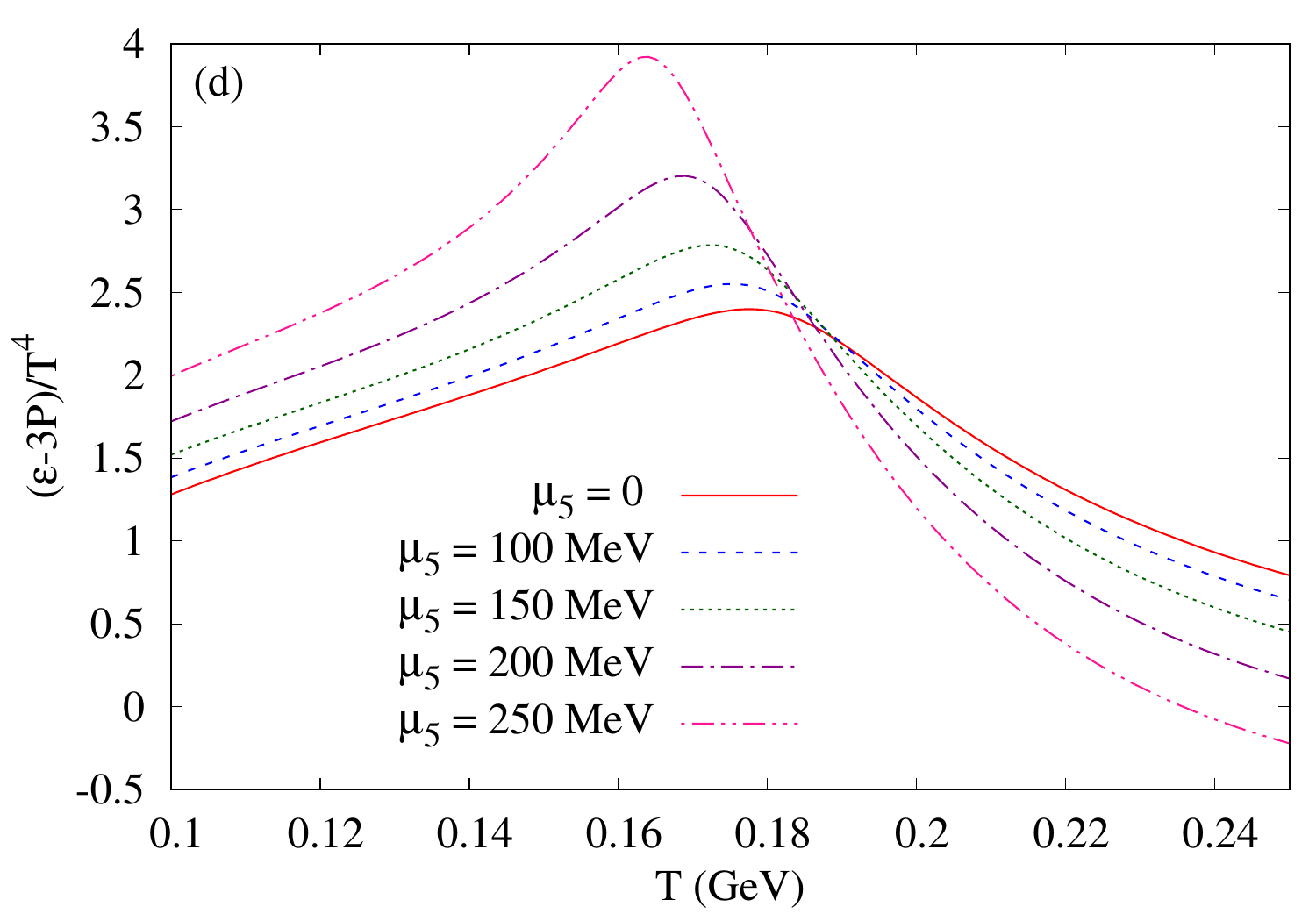}
		\caption{(Color Online) The variation of dimensionless-scaled (a) energy density $\dfrac{\varepsilon}{T^4}$, (b) pressure $\dfrac{3P}{T^4}$, (c) entropy density $\dfrac{3s}{4T^3}$, and (d) trace anomaly $\dfrac{\varepsilon-3P}{T^4}$ as a function of temperature at different values of CCP.}
		\label{fig.thermodynamics}
	\end{center}
\end{figure}
\begin{figure}[h]
	\begin{center}
		\includegraphics[angle=-0,scale=0.35]{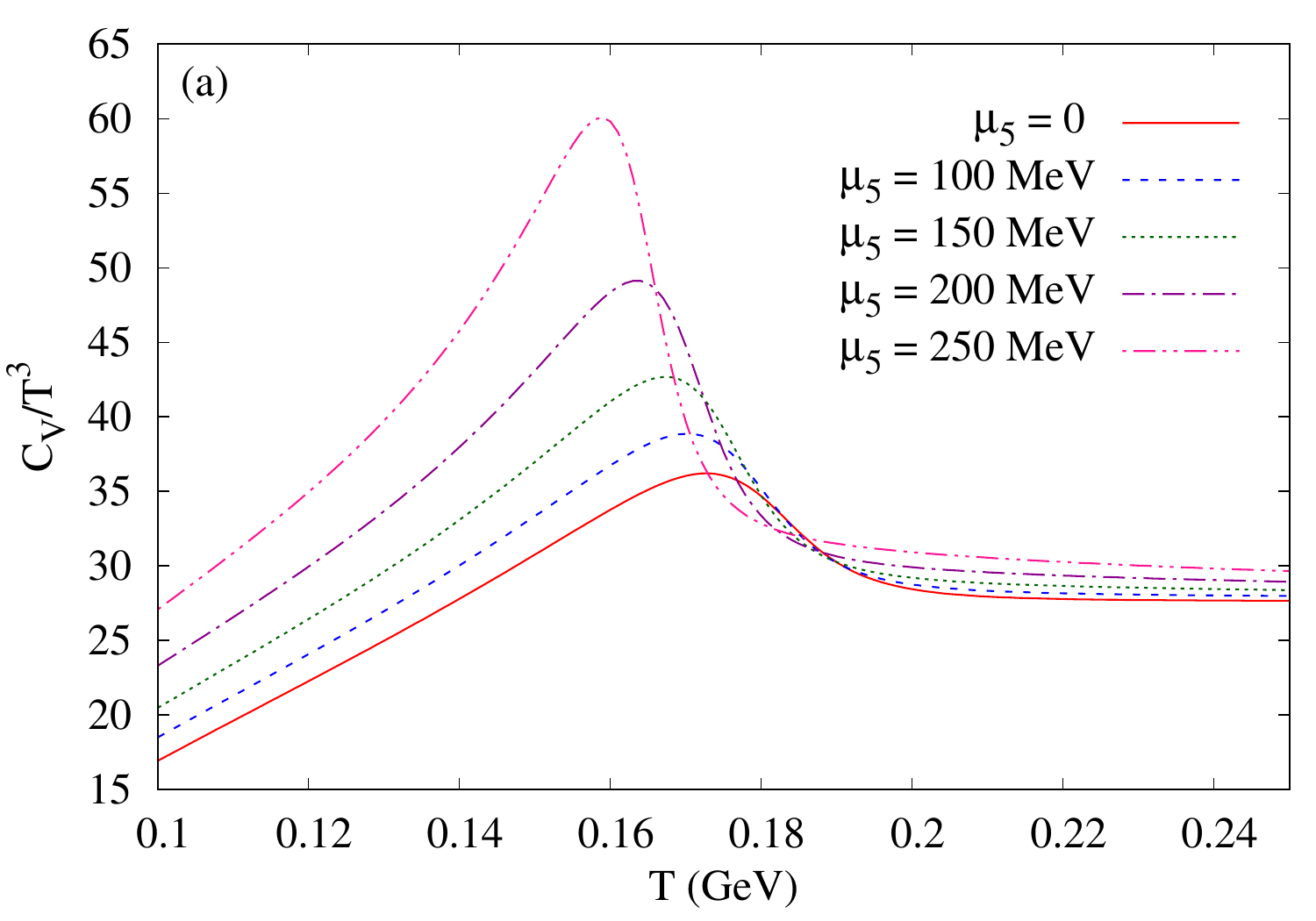} \includegraphics[angle=-0,scale=0.35]{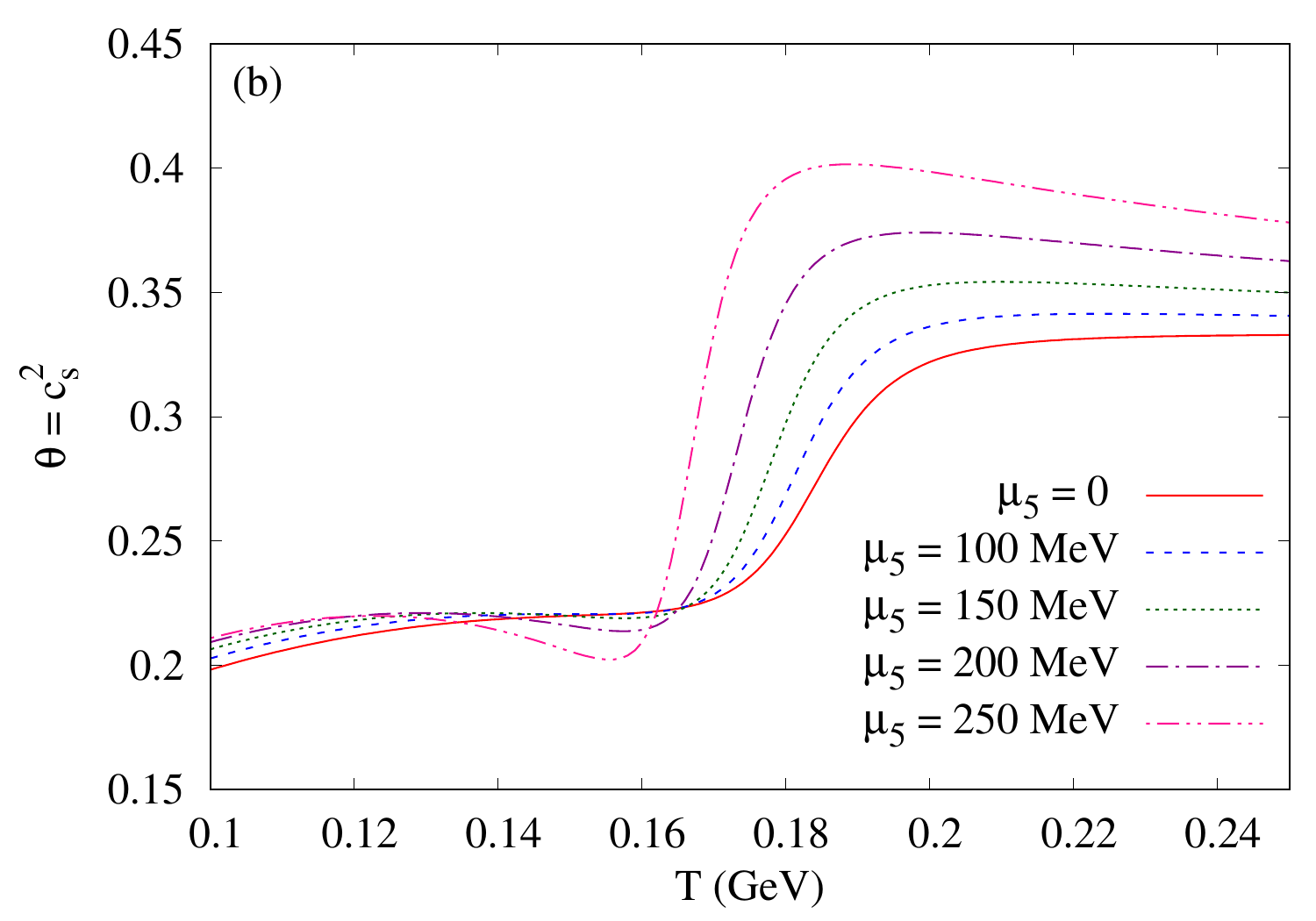} 
		\caption{(Color Online) The variation of (a) dimensionless-scaled specific heat $\dfrac{C_V}{T^3}$, and (b) isentropic squared speed of sound  $\theta=c_s^2$ as a function of temperature at different values of CCP.}
		\label{fig.cv.theta}
	\end{center}
\end{figure}

We now move to the numerical results of the thermodynamic quantities. In Figs.~\ref{fig.thermodynamics}(a), (b), (c) and (d) we have respectively shown the variation of dimensionless-scaled energy density $\dfrac{\varepsilon}{T^4}$, pressure $\dfrac{3P}{T^4}$, entropy density $\dfrac{3s}{4T^3}$, and trace anomaly $\dfrac{\varepsilon-3P}{T^4}$ as a function of temperature at different values of CCP. We first note that thermodynamic quantities are continuous near the transition temperature indicating a smooth cross-over type phase transition. All the quantities $\varepsilon$, $P$ and $s$ increase monotonically with increasing temperature up to the transition temperature, after which they show a saturation type behavior at high temperature approaching their respective Stefan-Boltzmann (SB) limit; thus qualitatively mimicking the typical QCD thermodynamics. The behavior of these quantities in the symmetry restored phase (in high temperature region) can be explained by noting that, in the massless limit, the thermodynamic quantities have the following analytic expressions at finite CCP:
\begin{eqnarray}
	P(M=0) &=& \frac{N_cN_f}{6} T^4 \TB{ \frac{7\pi^2}{30} + \FB{\frac{\mu_5}{T}}^2 }, \\
	s(M=0) &=& \frac{N_cN_f}{3} T^3 \TB{ \frac{7\pi^2}{15} + \FB{\frac{\mu_5}{T}}^2 }, \\
	\varepsilon(M=0) &=& \frac{N_cN_f}{6} T^4 \TB{ \frac{7\pi^2}{10} + \FB{\frac{\mu_5}{T}}^2 }.
\end{eqnarray}
Thus at high temperature, $\varepsilon$, $P$ and $s$ increase with the increase in CCP as can be observed in Fig.~\ref{fig.thermodynamics}(a)-(c). Moreover the temperature dependencies of scaled energy density, entropy density and trace anomaly exhibit local maxima at the transition temperature which moves toward lower value of temperature as we increase CCP as a consequence of ICC effect discussed earlier and can be clearly seen in Figs.~\ref{fig.thermodynamics}(a), (c) and (d).

Next in Fig.~\ref{fig.cv.theta}(a), we have depicted the variation of dimensionless-scaled specific heat $\frac{C_V}{T^3}$ as a function of temperature at different values of CCP. We notice that, in low $T$-region, $\frac{C_V}{T^3}$ increases with the increase in temperature reaching a maxima at transition temperature. Beyond the transition temperature $\frac{C_V}{T^3}$ decreases rapidly and shows a saturation behavior at high temperature, asymptotically approaching the SB limit. Except the transition region, $\frac{C_V}{T^3}$ increases with the increase in CCP; whereas near the transition temperature, a nonmonotonic CCP dependence is observed for specific heat. Moreover, the peak of the $\frac{C_V}{T^3}$ moves toward lower values of temperature with the increase in CCP owing to the ICC effect as can be observed in the figure.

Next in Fig.~\ref{fig.cv.theta}(b), we have shown the variation of isentropic squared speed of sound  $\theta=c_s^2$ as a function of temperature at different values of CCP. As $\theta = \frac{s}{C_V}$, the behavior of $\theta$ can be explained from Fig.~\ref{fig.thermodynamics}(c) and \ref{fig.cv.theta}(a). The maxima of $C_V$ at the transition temperature translates to a lower value of $\theta$ at the transition temperature which is visible in \ref{fig.cv.theta}(b). After the transition temperature $\theta$ increases rapidly in the symmetry restored phase followed by a saturation type behavior in the high $T$-region approaching the SB limit. With the increase in CCP, $\theta$ also increases for most of the temperature regions, except a small portion around the transition temperature. At finite CCP, we also observe that $\theta$ becomes more than $\frac{1}{3}$ in the symmetry restored phase thus crosses the conformal limit of $\mu_5=0$.
\begin{figure}[h]
	\begin{center}
		\includegraphics[angle=-0,scale=0.35]{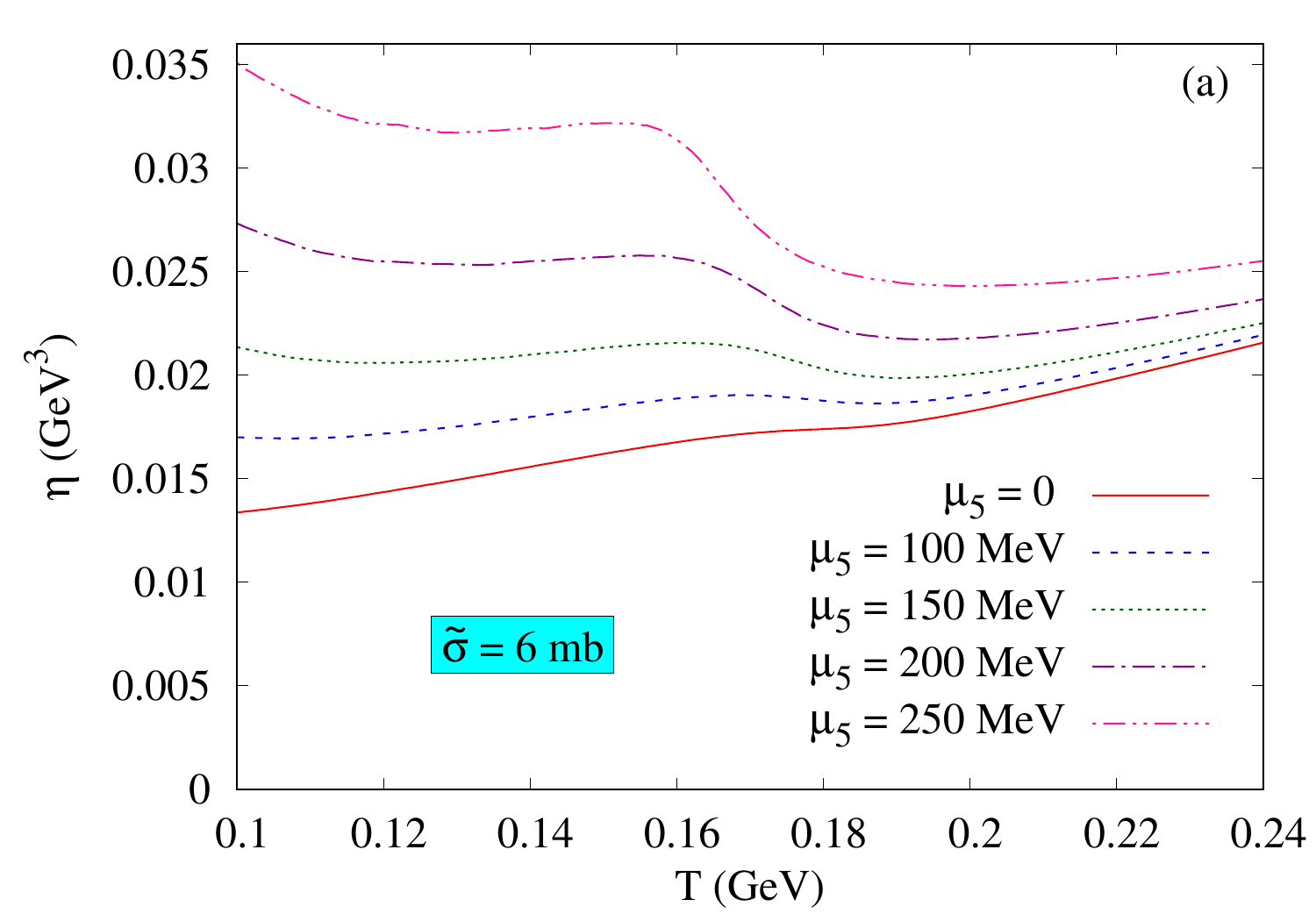} 
		\includegraphics[angle=-0,scale=0.35]{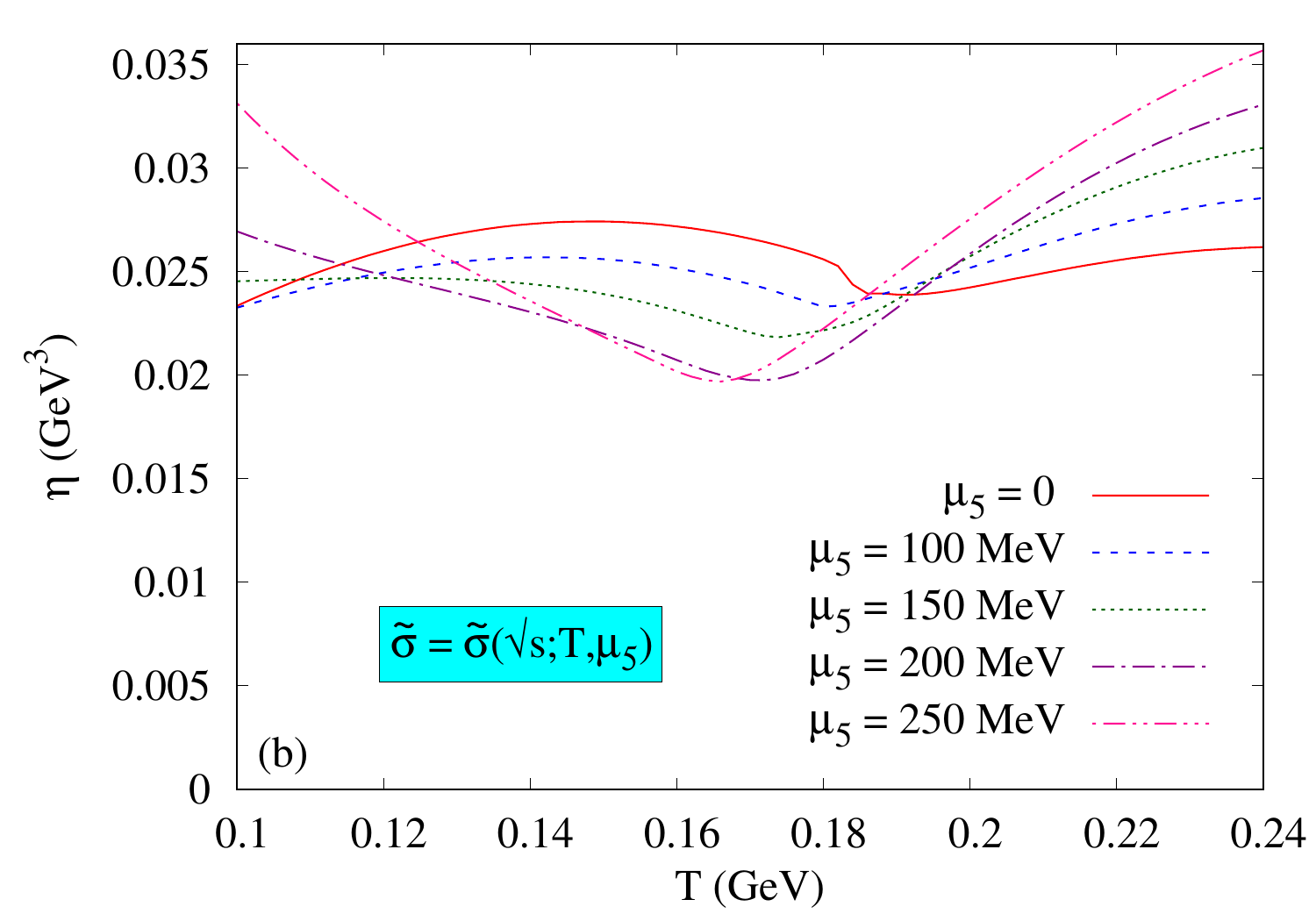} 
		\includegraphics[angle=-0,scale=0.35]{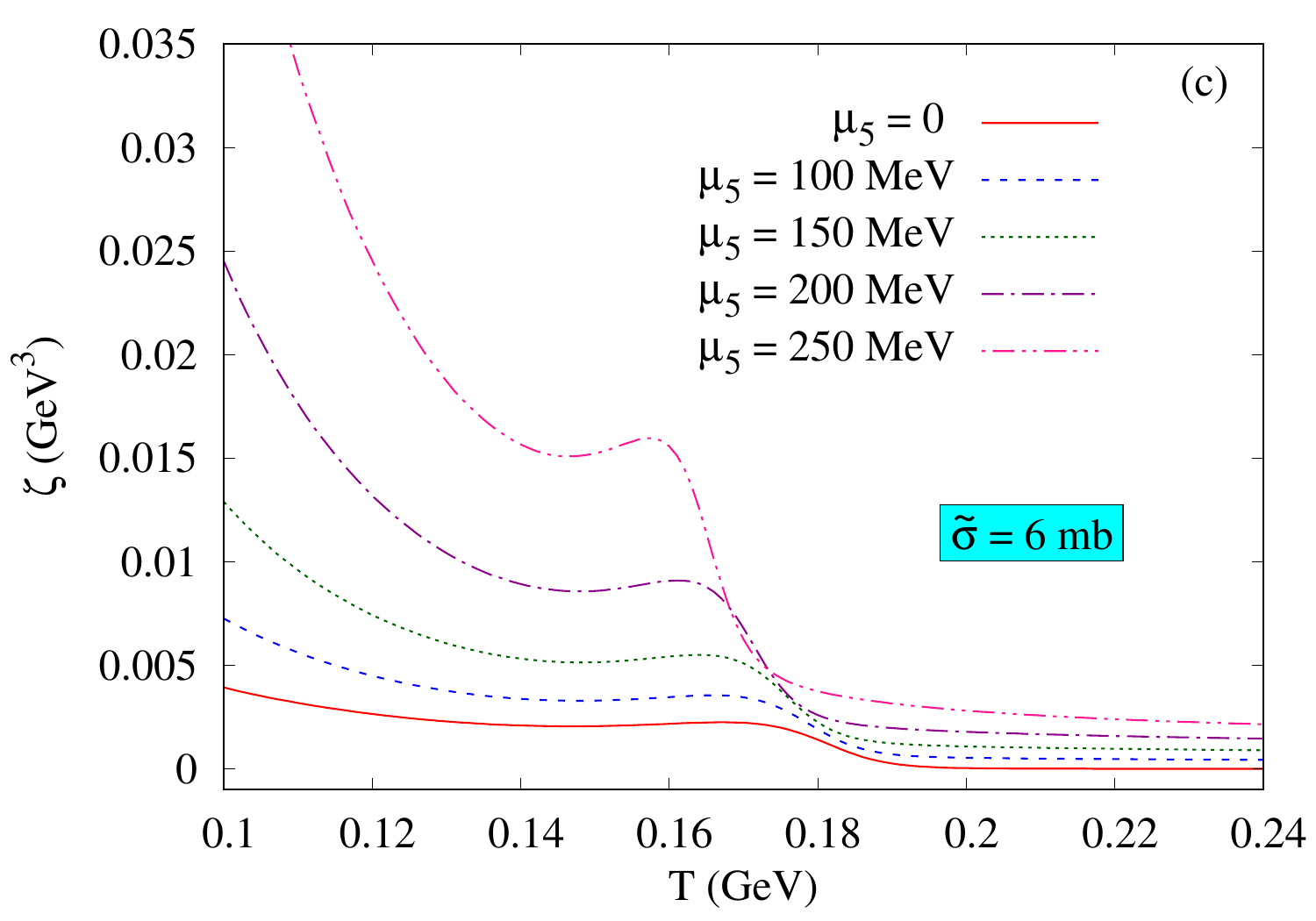}
		\includegraphics[angle=-0,scale=0.35]{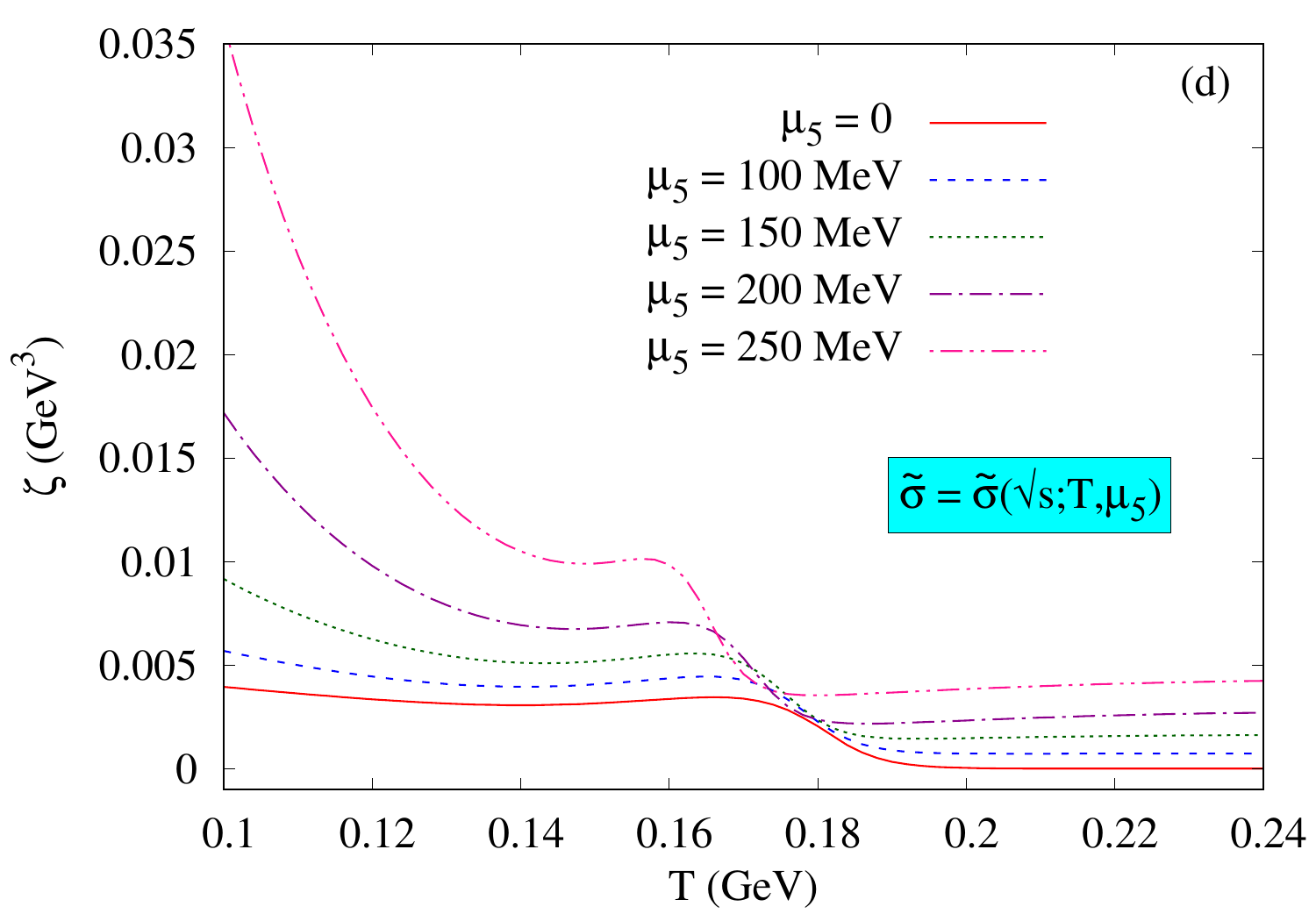}
		\caption{(Color Online) The variation of shear viscosity $\eta$ as a function of temperature for different values of CCP calculated (a) using a constant $\widetilde{\sigma} = 6$ mb, and (b) using $\widetilde{\sigma} = \widetilde{\sigma}(\mans;T,\mu_5)$. The variation of bulk viscosity $\zeta$ as a function of temperature at different values of CCP calculated (c) using a constant $\widetilde{\sigma} = 6$ mb, and (d) using $\widetilde{\sigma} = \widetilde{\sigma}(\mans;T,\mu_5)$.}
		\label{fig.viscosity}
	\end{center}
\end{figure}

We now come to the discussions of the numerical results of the temperature and CCP dependence of viscous coefficients. The expression of $\eta$ and $\zeta$ are given in Eqs.~\eqref{eta} and \eqref{zeta} both of which contain the dynamical quantity $\Gamma_r(\bm{k};T,\mu_5) = \tau_r^{-1}(\bm{k};T,\mu_5) \sim n \widetilde{\sigma}$, which in turn contain the scattering cross section $\widetilde{\sigma} = \widetilde{\sigma}(\mans;T,\mu_5)$. As the cross section $\widetilde{\sigma}$ has a mild $T$ and $\mu_5$ dependence as shown earlier in Fig.~\ref{fig.xsection}, the leading order temperature and CCP dependence of the viscous coefficients is expected to come from the other factors except $\widetilde{\sigma}(\mans;T,\mu_5)$ in Eqs.~\eqref{eta}, \eqref{zeta} and \eqref{gam}. To see this explicitly, we will first take a constant $\widetilde{\sigma}$ and study the temperature and CCP dependence of viscous coefficients; and later we will compare them with that of a full numerical calculation with $\widetilde{\sigma} = \widetilde{\sigma}(\mans;T,\mu_5)$.

In Fig.~\ref{fig.viscosity}(a), we have depicted the variation of shear viscosity $\eta$ as a function of temperature for different values of CCP calculated using a constant cross section $\widetilde{\sigma} = 6$ mb which is a typical value of the cross section as one can obtain from Fig.~\ref{fig.xsection}. For this constant cross section, $\eta$ is seen to increases slowly with the increase in $T$ at vanishing CCP. At finite values of CCP,  with the increase in temperature, $\eta$ first increases at even slower rate (than the zero-CCP case) in the low $T$-region, then drops suddenly at the phase transition temperature, and finally increases again.  The change in the transition temperature due to the ICC effect is also reflected in the $T$ dependence of $\eta$. On the other hand, $\eta$ increases with the increase in CCP for all the temperature ranges. A straightforward analysis of the $T$ and $\mu_5$ dependence of $\eta$ from Eq.~\eqref{eta} is difficult because of the significant $T$ and $\mu_5$ dependence of the constituent quark mass $M=M(T,\mu_5)$ which enters into different factors in  Eq.~\eqref{eta}. 

Next in Fig.~\ref{fig.viscosity}(b), we have shown the variation of shear viscosity $\eta$ as a function of temperature for different values of CCP calculated using a full energy/momentum, temperature and CCP-dependent cross section $\widetilde{\sigma} = \widetilde{\sigma}(\mans;T,\mu_5)$. In this case, we observe for all CCP values that, in most of the temperature  region, $\eta$ first decreases, attains a local minima at the transition temperature, and finally increases with the increase in temperature. For extreme low temperature $T\lesssim 120$ MeV and CCP $\mu_5 \lesssim 150$ MeV, we notice additional nonmonotonicity in the $T$-dependence of $\eta$.  The position of these local minima moves toward lower value of $T$ with the increase in CCP due to the ICC effect as can be clearly noticed. The CCP dependence of $\eta$ at the high and extreme low temperature regions  ($200 \lesssim T \lesssim 120$ MeV) is found to be more or less similar to Fig.~\ref{fig.viscosity}(a); in particular $\eta$ increases with the increase in $\mu_5$.
In Fig.~\ref{fig.viscosity}(b), we observe a qualitative difference for the $T$ and $\mu_5$-dependence of $\eta$ from that of Fig.~\ref{fig.viscosity}(a) which is caused by nonlinear energy/momentum, temperature and CCP dependence of $\widetilde{\sigma}(\mans;T,\mu_5)$.

In order to understand the behavior of the shear viscosity in a semi-qualitative manner we proceed as follows. In the limit of constant $\sigma$ and using $\Gamma\sim n\sigma\ensembleaverage{v}$ with $\ensembleaverage{v}$ being the average thermal speed, the expression for $\eta$ is given by 
\begin{eqnarray}
\eta(T,\mu_5) \sim \frac{2N_cN_f}{15 n \sigma\ensembleaverage{v}} \cdot \frac{1}{T} \int\!\! \frac{d^3k}{(2\pi)^3} \sum_{r \in \{\pm\}}  \frac{1}{(\wkr)^2} \TB{k^2 (k+r\mu_5)^2}  \fkr\fb{1 -\fkr}, \label{eta.a}
\end{eqnarray}
which for massless Fermions can be expanded as
\begin{eqnarray}
\eta(T,\mu_5; M=0) \sim \frac{T}{\sigma}   \Big[ \frac{\zeta (5)}{ \zeta (3)} +\frac{6 \zeta (3)^2-5\zeta (5)  \log (4)}{15\zeta (3)^2} \FB{\frac{\mu_5}{T}}^2
+O\FB{\frac{\mu_5}{T}}^4 \Big]. \label{eta.a.2}
\end{eqnarray}
From this expression we expect a linear $T$ dependence of $\eta$ in the high temperature region which is visible in Figs.~\ref{fig.viscosity}(a) as well as (b) above. At high values of $\mu_5$, the lower temperature region of $\eta$ is expected to show non-monotonicity in its $T$-dependence due to the presence of the $\FB{\frac{\mu_5}{T}}$-terms. However, the actual non-monotonicity observed in the $T$-dependence of $\eta$ at high CCP seen in Figs.~\ref{fig.viscosity}(a) and (b) may be attributed to the competition of the $n=n(T,\mu_5)$ in the denominator and other factors of the numerator including the integral in Eq.~\eqref{eta.a}.

Let us now move on the discussions of the bulk viscosity.  In Fig.~\ref{fig.viscosity}(c), we have shown the variation of bulk viscosity $\zeta$ as a function of temperature for different values of CCP calculated using a constant cross section $\widetilde{\sigma} = 6$ mb. We first note that, at zero CCP $\zeta$ vanishes in the massless case (conformal system) as can be observed from Eq.~\eqref{zeta.0} in which $\theta$ becomes $\frac{1}{3}$ for $M=0$. Thus it is expected at zero CCP that the bulk viscosity will become zero in the chiral symmetry restored phase where $M \simeq 0$. The vanishing of $\zeta$ at high temperature at zero CCP is clearly noticeable in solid-red curve in Fig.~\ref{fig.viscosity}(c). However, at finite CCP, the bulk viscosity does not vanish in the conformal limit due to the additional $\mu_5$-dependent terms within the square bracket of Eq.~\eqref{zeta} even if $\theta$ is put to $\frac{1}{3}$ by hand. This behavior can be observed in  Fig.~\ref{fig.viscosity}(c) by looking at the non-zero CCP curves at high $T$-region which are well separated from zero. Moreover, for all the values of CCP, we find that $\zeta$ slowly decreases with the increase in $T$ in the low $T$-region, then it suffers a sudden decrease at the transition temperature and finally shows a saturating behavior in the high $T$-region. The position of the transition temperature (where the sudden jump of $\zeta$ occurs) on the $T$-axis is also seen to move toward the lower value of temperature with the increase in CCP due to ICC effect. In this case, the significant $T$ and $\mu_5$ dependence of $\theta(T,\mu_5)$ also contribute to the $T$ and $\mu_5$ dependence of $\zeta$ which was not present in case of $\eta$.

Next, in Fig.~\ref{fig.viscosity}(d), we have shown the variation of bulk viscosity $\zeta$ as a function of temperature for different values of CCP calculated using a full energy/momentum, temperature and CCP-dependent cross section $\widetilde{\sigma} = \widetilde{\sigma}(\mans;T,\mu_5)$. The qualitative nature of the curves in Fig.~\ref{fig.viscosity}(d) is exactly identical to that of Fig.~\ref{fig.viscosity}(c) except the fact that $\zeta$ increases at a larger (slower) rate with the increase in CCP in the high (low) temperature region as compared to  Fig.~\ref{fig.viscosity}(c). Thus the energy/momentum, temperature and CCP-dependence of the cross section has an insignificant effect on the qualitative behavior of bulk viscosity unlike the case of shear viscosity. 
\begin{figure}[h]
	\begin{center}
		\includegraphics[angle=-0,scale=0.35]{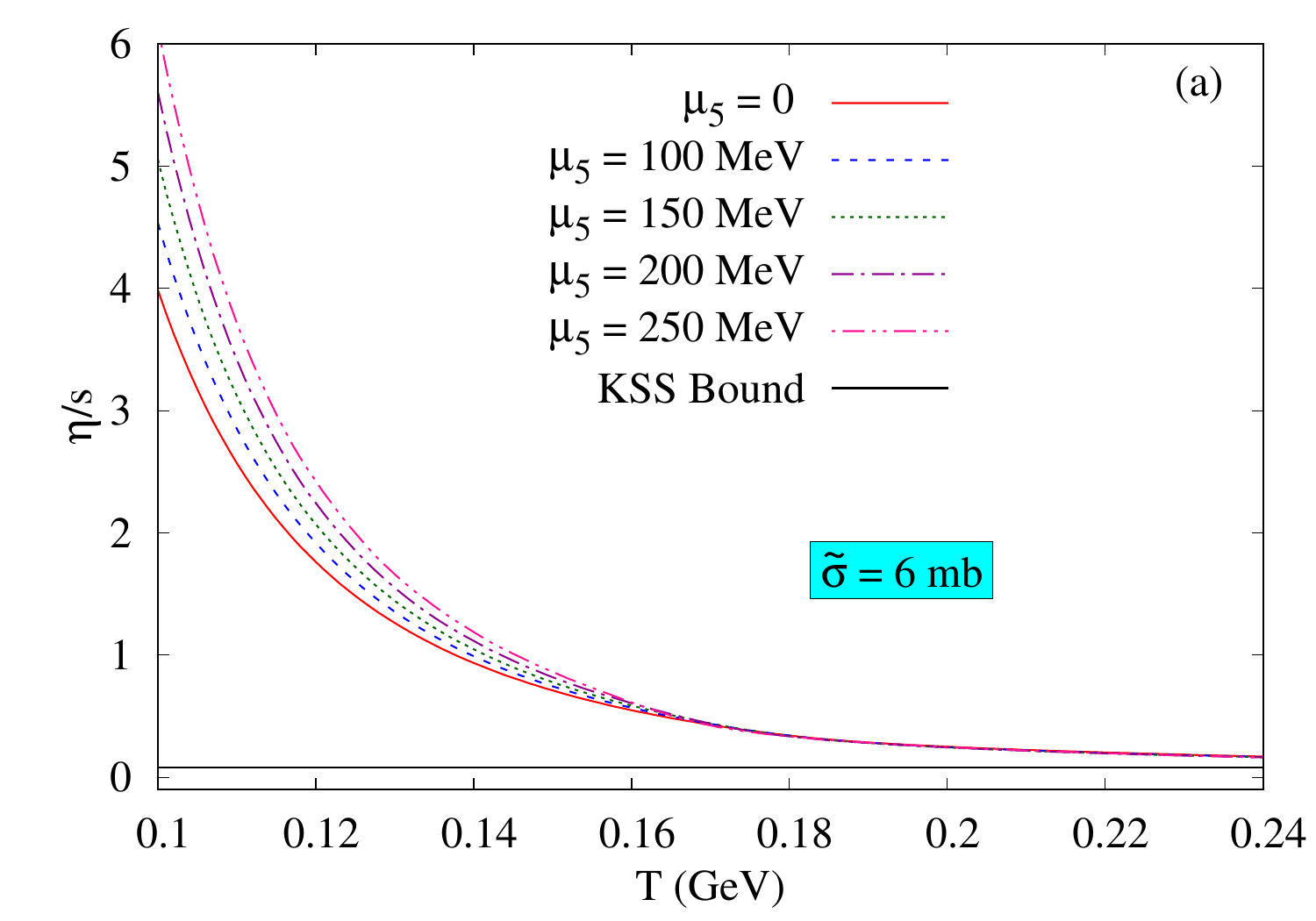}
		\includegraphics[angle=-0,scale=0.35]{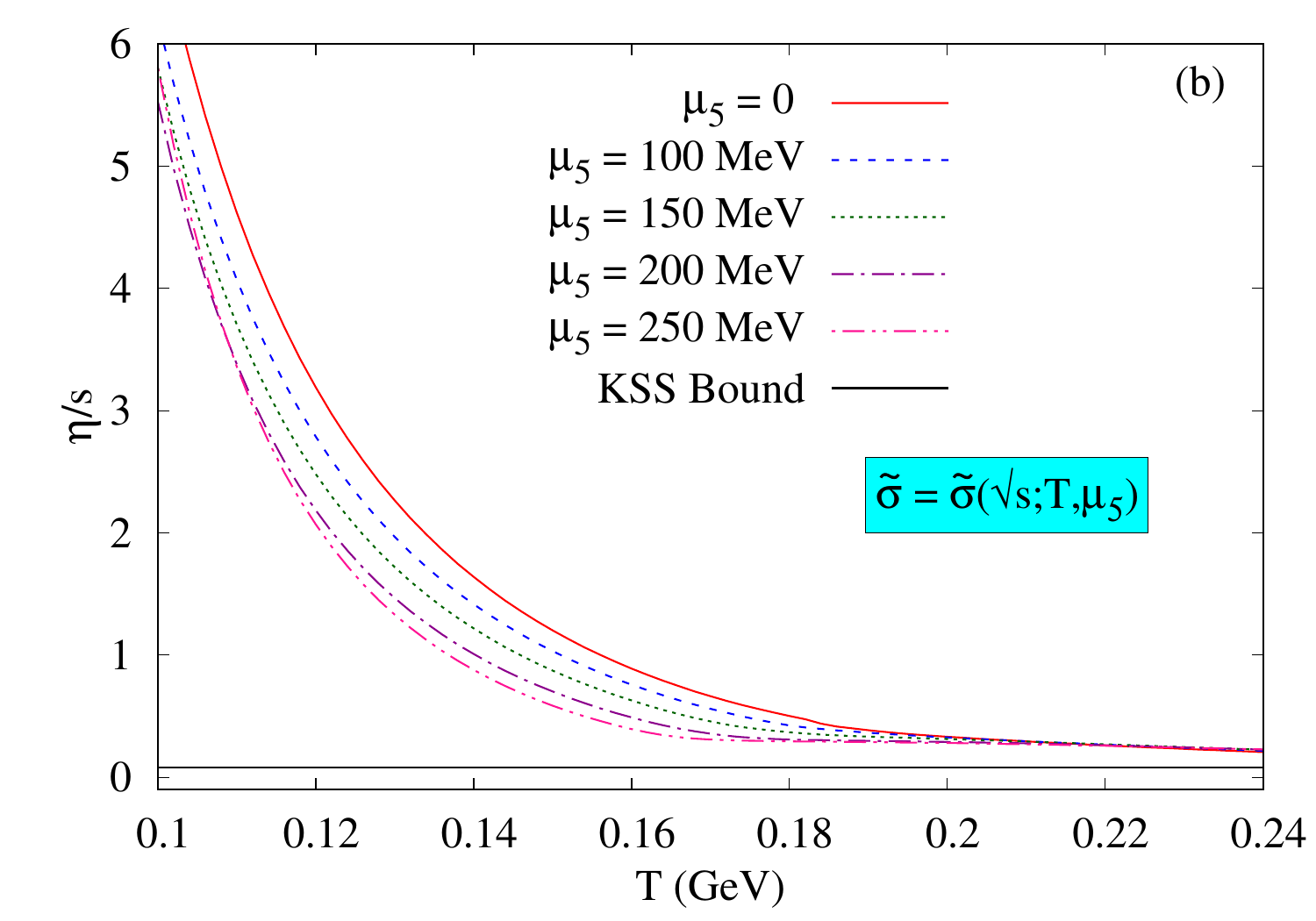}
		\includegraphics[angle=-0,scale=0.35]{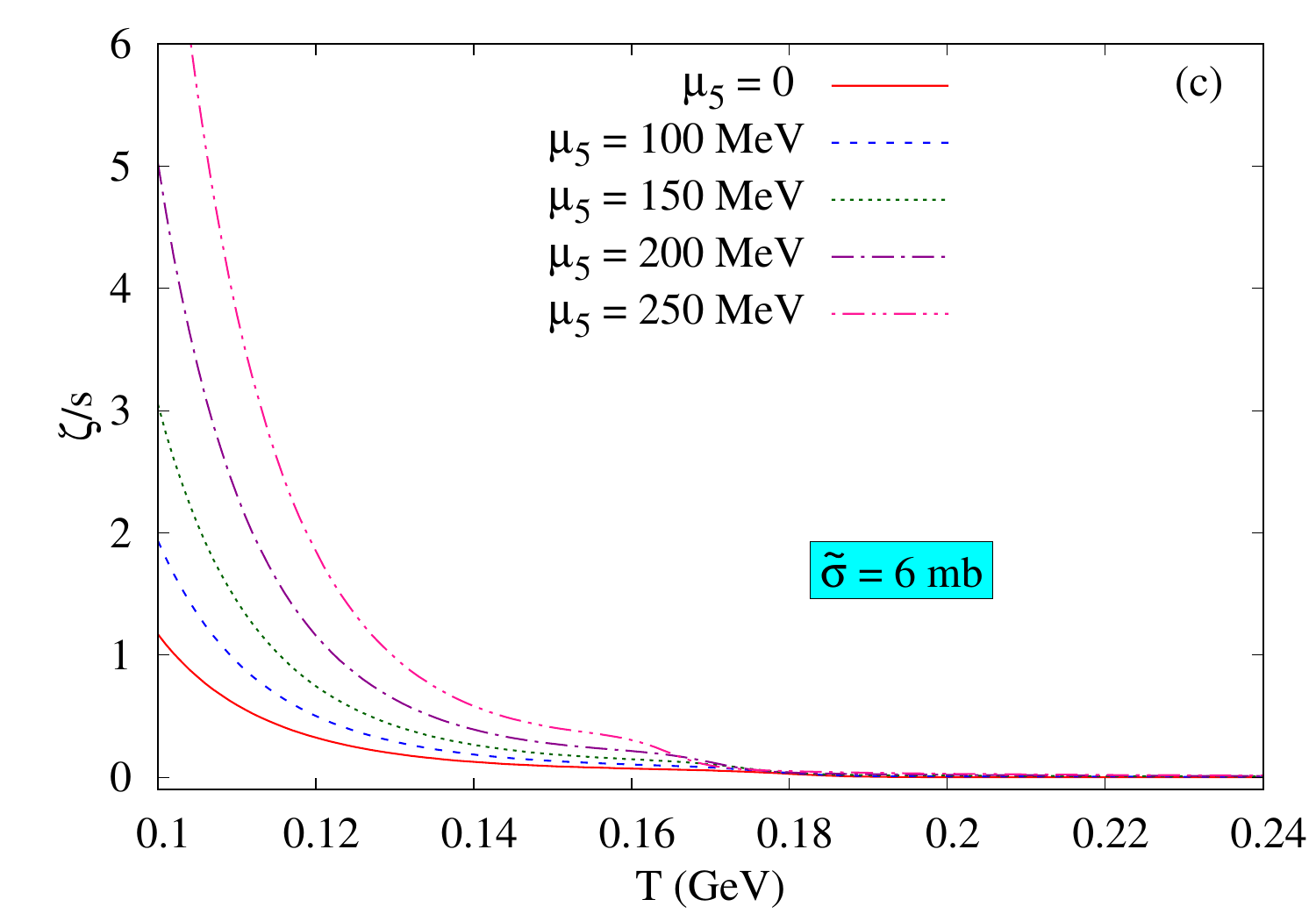}
		\includegraphics[angle=-0,scale=0.35]{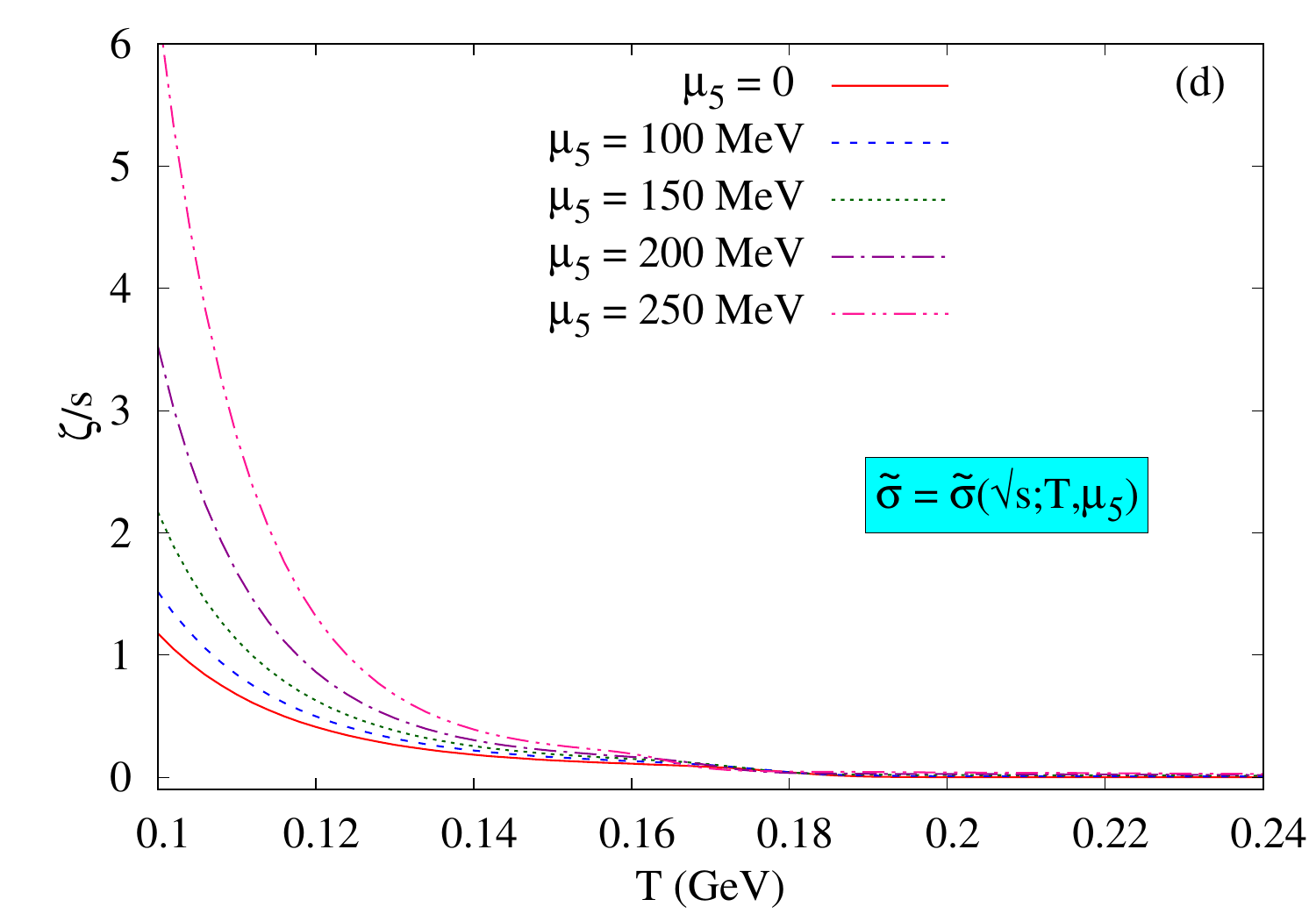}
	\caption{(Color Online) The variation of shear viscosity  to entropy density ratio $\eta/s$ as a function of temperature for different values of CCP calculated (a) using a constant $\widetilde{\sigma} = 6$ mb, and (b) using $\widetilde{\sigma} = \widetilde{\sigma}(\mans;T,\mu_5)$. The variation of bulk viscosity  to entropy density ratio $\zeta/s$ as a function of temperature at different values of CCP calculated (c) using a constant $\widetilde{\sigma} = 6$ mb, and (d) using $\widetilde{\sigma} = \widetilde{\sigma}(\mans;T,\mu_5)$. The KSS limit $1/4\pi$ is also shown in subfigures (a) and (b) by solid-black line}
	\label{fig.viscositybys}
	\end{center}
\end{figure}

We finally come to the discussions of the temperature and CCP dependence of viscosity to entropy density ratios $\eta/s$ and $\zeta/s$ which are also called specific shear and bulk viscous coefficients.  In Fig.~\ref{fig.viscositybys}(a), we have shown the variation of $\eta/s$ as a function of temperature for different values of CCP calculated using a constant cross section $\widetilde{\sigma} = 6$ mb. We notice that, the temperature dependence of $\eta/s$ is mainly controlled by the entropy density $s\sim T^3$ which makes the $\eta/s$ to decreases rapidly with the increase in temperature. The CCP dependence of $\eta/s$ results from the competition between the CCP dependence of $\eta$ and $s$ both of which increases with the increase in CCP as can be seen in Fig.~\ref{fig.viscosity}(a) and \ref{fig.thermodynamics}(c). However, in our case $\eta$ wins over $s$ and the CCP dependence of $\eta/s$ becomes similar to the CCP dependence of $\eta$. In particular, $\eta/s$ is found to increase significantly in the low temperature region with the increase in CCP.  However at high temperature, the CCP effects are washed out and the curves merge with each other. The $\eta/s$ at high temperature becomes close to the KSS limit $\eta/s = \frac{1}{4\pi}$ which has been shown by solid-black line in Fig.~\ref{fig.viscositybys}(a).

Next, in Fig.~\ref{fig.viscositybys}(b),  we have shown the variation of specific shear viscosity $\eta/s$ as a function of temperature for different values of CCP calculated using $\widetilde{\sigma} = \widetilde{\sigma}(\mans;T,\mu_5)$.  The temperature dependence of $\eta/s$ in this case is found to be identical with that of Fig.~\ref{fig.viscositybys}(a); whereas the CCP dependence become opposite to  Fig.~\ref{fig.viscositybys}(a). In particular, using  $\widetilde{\sigma}(\mans;T,\mu_5)$, we found that $\eta/s$ decreases with the increase in CCP for almost all CCP values. This is due to the fact that $\eta$ decreases with the increase in CCP for most of the temperature region whereas $s$ increases with the increase in CCP. In this case also, the $\eta/s$ becomes close to the KSS limit at high temperature, and the $\mu_5$-dependence is washed out alike  Fig.~\ref{fig.viscositybys}(a).

Next in  in Fig.~\ref{fig.viscositybys}(c), we have shown the variation of $\zeta/s$ as a function of temperature for different values of CCP calculated using $\widetilde{\sigma} = 6$ mb. For all values of CCP, $\zeta/s$ decreases monotonically with the increase in temperature. This is due to the fact that $\zeta$ decreases with the increase in $T$, whereas the entropy density $s\sim T^3$ increases with the increase in $T$. We notice that $\zeta/s$ also increases with the increase in CCP alike $\zeta$. Thus  the CCP dependence of $\zeta/s$ is dominated by the CCP dependence of $\zeta$ since both the $\zeta$ and $s$ increases with the increase in CCP as can be seen in Fig.~\ref{fig.viscosity}(c) and \ref{fig.thermodynamics}(c). At high temperature, the CCP dependence in $\zeta/s$ washed out alike $\eta/s$.

Finally,  in Fig.~\ref{fig.viscositybys}(d),  we have depicted the variation of specific bulk viscosity $\zeta/s$ as a function of temperature for different values of CCP calculated using the full energy/momentum, temperature and CCP-dependent cross section $\widetilde{\sigma} = \widetilde{\sigma}(\mans;T,\mu_5)$.  The temperature as well as CCP dependence of $\zeta/s$ in this case is found to be identical with that of Fig.~\ref{fig.viscositybys}(c).  However, at high temperature we see slight CCP dependence in $\zeta/s$; in particular the different CCP curves are slightly separated even at $T=240$ MeV.

\section{Summary And Conclusion} \label{sec.summary}
In summary, we have evaluated the shear $\eta$ and bulk $\zeta$ viscous coefficients in a hot and chirally asymmetric quark matter characterized in terms of a chiral chemical potential (CCP) using the two-flavor Nambu-Jona--Lasinio (NJL) model. Following the Green-Kubo formalism we extract the coefficients from the long-wavelength limit of the in-medium spectral function corresponding to the energy momentum tensor (EMT) current correlator which is calculated using the real time formalism of finite temperature field theory. The momentum dependent thermal width of the quark/antiquark that enters into the expression of the viscosities as a dynamical input containing interactions, has been obtained from the $2\to2$ scattering processes mediated via the collective mesonic modes in scalar and pseudoscalar channels encoded in respective in-medium polarization functions having explicit temperature and CCP dependence. In addition, thermodynamical quantities such as pressure, energy density, entropy density $(s)$, specific heat and isentropic speed of sound have also been calculated at finite CCP. We also examine the phase diagram $T$ vs. $\mu_5$ which shows the ICC effect. 

We find that the isospin averaged $qq\to qq$ cross section shows Breit-Wigner like structure in its $\rs$ dependence whereas the same for $q\qbar\to q\qbar$ does not; this Breit-Wigner structure tends to disappear at high temperature owing to the mesonic spectral broadening. The total cross section for $qx\to qx$ is found to have significant (insignificant) temperature dependence in the low (high) CCP-region.  The momentum dependent quark/antiquark relaxation time $\tau_s(p;T,\mu_5)$ shows an overall decreasing trend with the increase in $p$; whereas its temperature and CCP dependence is found to be roughly $\tau_s \sim \frac{1}{T^3\SB{a+b \FB{\frac{\mu_5}{T}}^2 }}$. The thermodynamic quantities $\varepsilon$, $P$ and $s$ increases monotonically with the increase in temperature and CCP, and are continuous near the transition temperature indicating a smooth cross-over. The scaled trace anomaly $(\varepsilon-3P)/T^4$ and specific heat $C_V/T^3$  exhibit a maxima near the transition temperature. The isentropic squared speed of sound shows a sudden jump near the transition temperature followed by a saturation type behavior in the high $T$-region approaching the SB limit.

It is observed that the shear viscous coefficient $\eta$ first decreases with temperature at finite CCP, attains a local minimum at the transition temperature and then increases with temperature. With the increase in CCP, $\eta$ is found to increase at the high and extreme low temperature regions; and decrease at the moderate temperature region near the phase transition. On the other hand, the bulk viscous coefficient $\zeta$ slowly decreases with the increase in $T$ in the low $T$-region for a given CCP, then it suffers a sudden decrease at the transition temperature and finally shows a saturating behavior in the high $T$-region. The specific shear viscosity $\eta/s$ decreases monotonically with the increase in temperature as well as CCP in the low temperature regions; however at high temperature, the CCP effects are washed out. The $\eta/s$ at high temperature is found to become close to the KSS limit. On the other hand, the specific bulk viscosity $\zeta/s$ is found to decrease (increase) with the increase in temperature (CCP).

\section*{ACKNOWLEDGMENTS}
S.G. is funded by the Department of Higher Education, Government of West Bengal, India. N.C., S.S. and P.R. are funded by the Department of Atomic Energy (DAE), Government of India. 

\section*{DATA AVAILABILITY}
No data were created or analyzed in this study.


\bibliography{z-MS}

\end{document}